\documentclass[aps,noshowpacs,superscriptaddress,nofootinbib,preprint,11pt]{revtex4}
\usepackage{graphicx}
\usepackage{array}
\usepackage{placeins}

\usepackage{amsmath, amssymb, graphics, setspace, hyperref}
\newcommand{\beq}{\begin{equation}}
\newcommand{\eeq}{\end{equation}}
\newcommand{\nn}{\nonumber}

\newcommand{\mF}{\mathcal{F}}

\newcommand{\mH}{\mathcal{H}}
\newcommand{\mK}{\mathcal{K}}
\newcommand{\mL}{\mathcal{L}}
\newcommand{\mM}{\mathcal{M}}
\newcommand{\mN}{\mathcal{N}}
\newcommand{\mQ}{\mathcal{Q}}
\newcommand{\mS}{\mathcal{S}}
\newcommand{\mU}{\mathcal{U}}
\newcommand{\mV}{\mathcal{V}}

\newcommand{\p}{\partial}

\newcommand{\f}{\frac}

\newcommand{\al}{\alpha}
\newcommand{\be}{\beta}
\newcommand{\ga}{\gamma}         
\newcommand{\de}{\delta}        
\newcommand{\ep}{\epsilon}

\newcommand{\ze}{\zeta}
\newcommand{\et}{\eta}

\newcommand{\la}{\lambda}       \newcommand{\La}{\Lambda}

\newcommand{\rh}{\rho}

\newcommand{\si}{\sigma}         \newcommand{\Si}{\Sigma}

\newcommand{\ta}{\tau}
      
\newcommand{\ph}{\phi}          
\newcommand{\ps}{\psi} 

\newcommand{\ch}{\chi}
\newcommand{\om}{\omega}        \newcommand{\Om}{\Omega}

\newcommand{\he}{\hat{e}}
\newcommand{\na}{\nabla}

\newcommand{\ran}{\rangle}
\newcommand{\bz}{\bar{z}}
\newcommand{\bi}{\bar{i}}
\newcommand{\bj}{\bar{j}}

\newcommand{\bN}{\bar{N}}
\newcommand{\bL}{\bar{L}}
\newcommand{\bX}{\bar{X}}
\newcommand{\bF}{\bar{F}}
\newcommand{\bM}{\bar{M}}

\begin{document}
\title{On the Supersymmetry of Bianchi attractors in Gauged supergravity}
\author{Bidisha Chakrabarty}
\email{bidisha@iopb.res.in}
\affiliation{Institute of Physics, Sachivalaya Marg, Bhubaneswar 751005, Odisha, India}
\affiliation{Homi Bhabha National Institute, Training School Complex,
Anushakti Nagar, Mumbai 400085, India}
\author{Karthik Inbasekar}
\email{ikarthik@theory.tifr.res.in}
\affiliation{The Raymond and Beverly Sackler School of Physics and Astronomy,
Tel Aviv University, Ramat Aviv 69978, Israel}
\author{Rickmoy Samanta}
\email{rickmoysamanta@gmail.com}
\affiliation{Department of Physics, Bar Ilan University, Ramat Gan, 52900, Israel.}

\begin{abstract}
Bianchi attractors are near horizon geometries with homogeneous symmetries in the spatial
directions. We construct supersymmetric Bianchi attractors in $\mathcal{N}=2, d=4,5$ 
gauged supergravity. In $d=4$ we consider gauged supergravity coupled to vector and hypermultiplets. 
In $d=5$ we consider gauged supergravity coupled to vector multiplets with a generic gauging of symmetries
of the scalar manifold and the $U(1)_R$ gauging of the $R$-symmetry. Analyzing the gaugino conditions we show that 
when the fermionic shifts do not vanish
there are no supersymmetric Bianchi attractors. This is analogous to the known condition that for maximally supersymmetric solutions, all the fermionic shifts must vanish.
When the central charge satisfies an extremization
condition, some of the fermionic shifts vanish and supersymmetry requires that the symmetries of the
scalar manifold do not be gauged. This allows supersymmetric Bianchi attractors sourced by massless gauge
fields and a cosmological constant. In five dimensions in the Bianchi I class we show that the anisotropic 
$AdS_3\times\mathbb{R}^2$ solution is $1/2$ BPS. We also construct a new class of $1/2$ BPS Bianchi
III geometries labeled by the central charge. When the central charge takes a special value the
Bianchi III geometry reduces to the known $AdS_3\times\mathbb{H}^2$ solution. For the Bianchi V and
VII classes the radial spinor breaks all of supersymmetry. We briefly discuss the conditions for possible massive supersymmetric Bianchi solutions by generalizing the matter content to include tensor, hypermultiplets and a generic gauging on the R symmetry.
\end{abstract}
\vskip .5 true cm

\maketitle
\section{Introduction}

In recent years intensive research on extremal black holes in $AdS$ space have unveiled
relations between seemingly unrelated fields such as gravity and condensed matter systems.
In $AdS/CFT$ extremal black holes provide the bulk gravitational description of zero temperature
ground states in strongly coupled field theories \cite{Aharony:1999ti}. Many condensed matter
systems show novel and diverse phase structures. At the quantum critical point the field theory
description is strongly coupled and exhibit phase transitions at zero temperature due to
quantum fluctuations \cite{sachdev2007quantum,Sachdev:2011wg}. The presence of diverse phases in the
field theory predict an equally large number of dual extremal geometries in the bulk. It is an
interesting program to identify and classify various possible extremal geometries. Some of the
earlier work in this direction have identified extremal geometries that exhibit Lifshitz and
hyperscaling violations \cite{Goldstein:2009cv,Goldstein:2010aw,Taylor:2008tg,Kachru:2008yh,
Balasubramanian:2008dm,Perlmutter:2010qu}. 
Of more recent research interest are extremal black branes dual to field theories with reduced
symmetries \cite{Donos:2011qt,Donos:2011ff,Donos:2012gg,Iizuka:2012iv,Iizuka:2012pn,Donos:2012wi,
Cremonini:2012ir, Erdmenger:2013zaa,Iizuka:2014iva,Donos:2014oha,Kiritsis:2015oxa,Ammon:2016szz}. Some of these
examples are anisotropic and display interesting phenomena such as violation of the KSS bound
\cite{Kovtun:2004de} when the anisotropy becomes much larger than the temperature
\cite{Jain:2015txa}.

In five dimensions homogeneous anisotropic extremal black brane geometries have been constructed in
\cite{Iizuka:2012iv,Iizuka:2012pn}. The metrics display manifest homogeneous symmetries in three
spatial directions. It is well known that the Killing vectors that generate these symmetries form
algebras that are isomorphic to real Lie algebras in dimension three. These real Lie algebras have
been well studied and are well known through the Bianchi classification
\cite{1975classical,ryan1975homogeneous}. The five dimensional geometries that display manifest
homogeneous symmetries in three spatial directions are referred to as the ``Bianchi attractors''.
These near horizon geometries are exact solutions to Einstein-Maxwell theories with massive/massless
gauge fields and a cosmological constant. \footnote{The terminology attractor is used because the
horizon geometries solve the
field equation exactly. Interpolating numerical solutions have been constructed in 
\cite{Kachru:2013voa}, justifying the terminology. However analytic solutions are much harder to
find.}

Black holes in $\mN=2$ supergravity exhibit a phenomenon known as the attractor
mechanism \cite{Ferrara:1995ih,Strominger:1996kf,Bellucci:2007ds,Ferrara:2008hwa}. In a black hole
background moduli fields flow to fixed point values at the horizon irrespective of their asymptotic
values at spatial infinity. The fixed point values are determined entirely in terms of the charges
carried by the black hole. As a result the Bekenstein-Hawking entropy of the black hole is
determined in terms of its charges. Although initial studies have focused on supersymmetric black
holes, it has been realized that the attractor mechanism is a consequence of
extremality \cite{Ferrara:1997tw}. Subsequently the attractor mechanism is generalized
to non-supersymmetric extremal black holes \cite{Sen:2007qy,Goldstein:2005hq}.

In recent years enormous effort has gone into generalizing the attractor mechanism
to gauged supergravities \cite{Halmagyi:2013qoa,Inbasekar:2012sh,BarischDick:2012gj,Klemm:2012yg,
Barisch:2011ui,Dall'Agata:2010gj,Hristov:2010eu,Ceresole:2001wi}. Significant progress has been made
especially for dyonic $AdS_4$ black holes \cite{Cacciatori:2009iz}. Large N index
computations in the dual twisted mass deformed ABJM theory find perfect matching of the microstate
counting with the Bekenstein Hawking entropy of the black hole \cite{Benini:2015eyy,Benini:2016rke}.
It is interesting to ask if the attractor mechanism generalizes to black brane geometries in
$AdS$. In this light, the first step is to embed these geometries in supergravity in order to study
their properties such as supersymmetry and stability.                                   
                                                                         
Some steps in this direction have been taken \cite{Kachru:2011ps,Inbasekar:2012sh} and explicit
examples of Bianchi attractors in $\mN=2$ gauged supergravity are constructed. However it turns
out that the geometries are non-supersymmetric and are unstable under linearized fluctuations
unless certain conditions are satisfied \cite{Inbasekar:2013vra,Inbasekar:2014lra}. The conditions
are such that there must exist a critical point of the effective potential and the Hessian of the
effective potential evaluated at the solution must have positive eigenvalues. For non-supersymmetric
extremal black hole solutions the above two conditions are sufficient to guarantee a stable Bianchi
attractor in gauged supergravity. However supersymmetric solutions always satisfy these conditions
and guarantee stability. 

In this work we look for supersymmetric Bianchi attractor geometries in $\mN=2$ gauged
supergravity. As a warmup, we study $d=4$ gauged supergravity coupled to vector and hyper multiplets
with a generic gauging of the symmetries of the hyper K\"{a}hler manifold. In four dimensions the
homogeneous symmetries are along the two spatial directions and the corresponding Lie algebras are
of two types namely Bianchi I and Bianchi II. Bianchi I geometries such as $AdS_5$
\cite{Ceresole:2001wi} and $z=2$ Lifshitz solution \cite{Halmagyi:2011xh,Cassani:2011sv} are well known solutions 
in this theory. In the Bianchi I case, we construct the $AdS_2\times\mathbb{R}^2$ geometry
sourced by time like gauge fields. We analyze the Killing spinor equations and find that the radial
spinor preserves $1/4$ of the supersymmetry. The gaugino and hyperino conditions impose additional
relations on the parameters of the theory. In the Bianchi II case, a $\f{1}{8}$ BPS
$AdS_2\times\mathbb{H}^2$ solution sourced by magnetic fields has been found recently in
\cite{Chimento:2015rra}. We construct a $AdS_2\times\mathbb{H}^2$ solution sourced by time like
gauge fields and find that the radial spinor breaks all of the supersymmetry. The Bianchi I and Bianchi II classes we studied in four dimensions correspond to the symmetries of $\mathbb{R}^2$ and $\mathbb{H}^2$. These are the only possible Bianchi classes of metric that one can construct in 3+1 dimensions with homogeneous symmetries in two spatial directions. Of course there exist more general manifolds like $T^2$ \cite{Halmagyi:2013qoa,Erbin:2014hsa}, however they do not belong to the Bianchi class and we do not consider them in our analysis.

In $d=5$ there exist a richer class of Bianchi attractor geometries. We consider the $ \mN=2$
gauged supergravity coupled to vector multiplets with a generic gauging of both 
symmetries of the very special manifold and the $U(1)_R$ subgroup of the $SU(2)_R$ symmetry group. From the gaugino conditions we find that there are no supersymmetric Bianchi attractors when the fermionic shifts in the supersymmetry variations are non vanishing.\footnote{In gauged supergravity literature the supersymmetry variations in the gaugino and hyperino that are proportional to the gauge coupling constant are referred to as fermionic shifts.} This is in the same spirit as the general analysis for maximally supersymmetric solutions \cite{Ceresole:2000jd,Ceresole:2001wi}. This result holds for a generic gauging of the scalar manifold, is dependent on the choice of the gauge field configuration that sources the solution and is  independent of the functional form of the Killing spinor. The basic argument is that the constant part of the Killing spinor should be a simultaneous eigenspinor of commuting matrices that can appear in the gaugino conditions. We find that for the known gauge field configurations that generate Bianchi type solutions, this does not happen in general. Independently we show that the radial Killing spinor for such solutions breaks all of supersymmetry.

When the central charge $Z$ of the solution satisfies an extremization condition, some of the fermionic shifts in the gaugino variations vanish. This is a reasonable condition to impose for any plausible geometry that can be an attractor solution.
Given this condition, supersymmetry invariance then requires that the effective mass term at the attractor point vanish. \footnote{One way to possibly avoid this is to consider tensor multiplets, we comment on this in \S\ref{gens}.} This condition allows Bianchi attractor solutions sourced by massless gauge fields since at the attractor point the ``effective mass terms'' in gauged supergravity are proportional to $g^2$. There are no further conditions from the gaugino variations and hence the supersymmetry of the solutions are entirely determined by the Killing spinor equation that follows from the gravitino variation. It is crucial to observe that the Killing spinor equation depends only on the gauge coupling constant of the R symmetry gauging, hence it follows that the Killing spinor integrability conditions (see eq 31 of \cite{Inbasekar:2012sh}) do not depend on the gauging of the scalar manifold. 

We construct Bianchi solutions sourced by massless gauge fields and a cosmological constant in the Bianchi I, Bianchi III, Bianchi V and Bianchi VII classes. In the Bianchi I case we find the anisotropic $AdS_3\times\mathbb{R}^2$ geometry recently
studied in \cite{Jain:2015txa,Critelli:2014kra} to be $1/2$ BPS. We also construct a supersymmetric
class of $1/2$ BPS Bianchi III geometries labeled by the central charge $Z$. When the central charge
of the solution takes special values, the geometry reduces to the known
$AdS_3\times\mathbb{H}^2$ \cite{Klemm:2000nj}. The Killing spinors in both these cases come in pairs
where one spinor is purely radial and the other spinor depends on both radial and transverse
coordinates other than $\mathbb{R}^2/\mathbb{H}^2$ directions. Moreover the constant part of the
spinors are eigenspinors of the radial Dirac matrix in all the above cases. For the Bianchi V and
Bianchi VII classes we find that the radial spinor breaks all supersymmetry. 

Finally, the presence of hyper and tensor multiplets can allow for some massive Bianchi attractor solutions in some special cases. In particular our results from the gaugino and Killing spinor conditions for the non-supersymmetric cases continues to hold even after including hypermultiplets and $SU(2)_R$ gauging as the Killing spinor equation is not affected seriously by this addition. However, addition of tensor multiplets will affect the analysis and depends crucially on the tensor field configuration in addition to new gaugino and hyperino conditions. We comment on the possibilities in \S\ref{gens} leaving a detailed analysis for future work.

The paper is organized as follows. In \S\ref{bianchidescription} we briefly describe homogeneous
symmetries and motivate Bianchi attractors. Following this we present the analysis for the $d=4$
Bianchi attractors in $\mN=2$ gauged supergravity in \S\ref{4dgs}. We move on to the five
dimensional case in \S\ref{5dgs}. In subsection \S\ref{gaugino5d} we present our main argument for
the absence of massive supersymmetric Bianchi attractors in gauged supergravity with $U(1)_R$ gauging and gauging of the symmetries of the very special manifold. Subsequently we analyze the Killing spinor equations for the massless cases in \S\ref{Ksmless}. In \S\ref{gens} we comment on the possible generalizations and necessary conditions when hyper and tensor multiplets are included with generic gauging. We present our conclusions and summarize in\S\ref{sumry}. In appendix \S\ref{conv} we provide useful supplementary material on spinors in $d=4,5$ and summarize our conventions. In appendices \S\ref{4ds} and \S\ref{new5d} we provide details about Bianchi type solutions in $d=4$ and $d=5$ respectively. The details of the Killing spinor equations for the massive cases are given in appendix \S\ref{masc}. 

\section{Homogeneous symmetries and Bianchi attractors}\label{bianchidescription}
In this section  we describe the homogeneous symmetries in two and three dimensions classified by
the Bianchi classification of Lie algebras. Towards the end we describe the ``Bianchi attractors''.
These are proposed near horizon geometries of extremal black branes with homogeneous symmetries in
the spatial directions \cite{Iizuka:2012iv}.

Consider a manifold $M$ endowed with a metric $g_{\mu\nu}$ that is invariant under a given set of
isometries. The Killing vectors $X_i$ that generate the isometries close to form an algebra
\beq
[X_i, X_j]=C_{ij}^{\ \ k}X_k
\eeq
where $C_{ij}^k$ are structure constants and they obey the usual Jacobi identity. The symmetry
group of the manifold is isomorphic to an abstract Lie group $G$ whose Lie algebra is generated by
the algebra of Killing vectors. 

A homogeneous manifold has identical metric properties at all points in space. Any two points on a
homogeneous space are connected by a symmetry transformation.  The symmetry group of a homogeneous
space of dimension $d$ is isomorphic to the group corresponding to $d$ dimensional real Lie algebra
\cite{ryan1975homogeneous,1975classical}.  On the other hand given the real Lie algebra in a
dimension $d$, it is possible to write the corresponding metric with manifest homogeneous
symmetries as follows. First one finds a basis of invariant vectors $e_i$ that commute with the
Killing vectors $X_i$
\beq
[X_i, e_i]=0 \ ,
\eeq
then the metric with homogeneous symmetries can be expressed in terms of one forms $\om^i$ dual to
the invariant vectors $e_i$ as
\beq\label{metinv}
ds^2 =g_{ij} \om^i\otimes \om^j
\eeq 
where $g_{ij}$ are constants. The invariant one forms satisfy the relation
\beq
d\om^k=\f{1}{2}C_{ij}^{\ \ k}\om^i \wedge \om^j
\eeq
where $C_{ij}^{\ \ k}$ are the same structure constants that appear in the algebra of the Killing
vectors. The real Lie algebras of dimension three fall into nine classes and are given by the well
known Bianchi classification. The structure constants and invariant one forms are listed in detail
in \cite{ryan1975homogeneous} (or see Appendix A of \cite{Iizuka:2012iv}).

As an illustration let us consider Bianchi VII, it has the symmetries of a three dimensional
euclidean space with translational symmetry along the $x$ direction and rotational symmetry along
the $(y,z)$ plane. The Killing vectors have the form
\beq
X_1 =\p_y\  , \ X_2 =\p_z \  , \ X_3 = \p_x+ y \p_z - z\p_y \ .
\eeq
These vectors close to form the Lie algebra
\beq
[X_1, X_2]=0 \  , \ [X_1,X_3]=X_2 \ , \ [X_2,X_3]=- X_1 \ .
\eeq
The structure constants are independent of spacetime coordinates. A nice way to see the
homogeneous symmetries manifest is to construct invariant vector fields $e_i$ that
commute with the Killing vectors
\beq
e_1 = \cos(x)\p_y + \sin(x)\p_z \ , \ e_2 = -\sin(x)\p_y + \cos(x)\p_z \ , e_3 =\p_x\ .
\eeq
The invariant one forms 
\beq
\om^1 =\cos(x)dy + \sin(x)dz \  , \ \om^2 = -\sin(x)dy + \cos(x)dz \ , \om^3 =dx 
\eeq
satisfy the algebra
\beq
d\om^1=-\om^2\wedge\om^3\ , \ d\om^2=\om^1\wedge\om^3 \ , \ d\om^3=0\ .
\eeq
The three dimensional euclidean metric invariant under the Bianchi VII symmetry can be written in
the form \eqref{metinv}. For example, in a diagonal basis as
\beq
ds^2 = (\om^1)^2+ (\om^2)^2+ \la^2(\om^3)^2
\eeq
where $\la$ is a constant. For the case where $\la=1$ the symmetry is enhanced to the usual
translation and rotational symmetries in the $(x,y,z)$ directions. Note that the Bianchi I and
Bianchi VII are sub algebras of the Poincar\'{e} algebra (see section \S \ref{algebrasx}).

To complete this discussion, it is useful to digress on the classification of two dimensional real
Lie algebras. These are classified in an analogous classification and are of precisely two types.
One is the simple Bianchi I algebra where all the generators commute. This is for instance the
symmetry group of $\mathbb{R}^2$. Obviously the Bianchi I algebra in dimension $2$ is a sub algebra
of the Bianchi I algebra in dimension $3$. While the only other non-trivial Bianchi algebra is the
one corresponding to the symmetries of the manifold $\mathbb{H} ^2$. The algebra
\beq
 [X_1,X_2]=X_1
\eeq
 is generated by the Killing vectors $X_i$, $i=1,2$,
\beq
X_1=\p_y\  , \ X_2 = \p_x + y \p_y \  .
\eeq
The invariant vector fields $e_i$ that commute with the Killing vectors are defined as
\beq
e_1= e^x \p_y \  , \ e_2=\p_x  \ , \  [e_1,e_2]=-e_1\ .
\eeq
The duals to the invariant one forms are given by
\beq
\om^1= e^{-x} dy\ , \om^2=dx \ , \ d\om^1= \om^1\wedge \om^2 \ .
\eeq
The metric invariant under this symmetry group takes the from
\beq
ds^2=  (\om^1)^2+(\om^2)^2\ .
\eeq 
Using the coordinate transformation $x=\ln \rh$ it is easy to see that this is precisely the metric
of Euclidean $AdS_2$. Note that the Bianchi II algebra corresponding to $\mathbb{H}^2$ in two
dimensions is a sub algebra of the Bianchi III algebra in three dimensions \cite{Inbasekar:2014lra}.

In this work we investigate the supersymmetry conditions on various Bianchi attractor
geometries described in \cite{Iizuka:2012iv}. The geometries have the general structure
\beq
ds^2 = -g_{tt}(r) dt^2 + g_{ij}(r) d\om^i \wedge d\om^j +dr^2
\eeq
where $i=1,2$ in case of four dimensions and the $\om^i$ are invariant one forms corresponding to
the homogeneous symmetries of two dimensional real Lie algebras described above. In
five dimensions $i=1,2,3$ and the corresponding $\om^i$ are invariant one forms given by the usual
Bianchi classification. The functions $g_{tt}(r)$ and  $g_{ij}(r)$ have a general form $e^{\be r}$,
where $\be$ are positive exponents. These metrics can be constructed as solutions to
Einstein-Maxwell theories with massive/massless gauge fields and a cosmological constant. As long as
the matter stress-tensor preserves the symmetries of the metric, explicit solutions can be
constructed for a wide range of parameters of the theory of interest. 

\section{Bianchi attractors in $\mN=2,d=4$ gauged supergravity}\label{4dgs}
In this section we describe $\mN=2, d=4$ gauged supergravity with $n_V$ vector and $n_H$ hyper
multiplets. We use the notations and conventions of \cite{Halmagyi:2011xh}, the relevant
conventions are summarized in appendix \ref{4dspinors}. The gravity multiplet consists of a metric
$g_{\mu\nu}$, a graviphoton $A_\mu^0$ and an $SU(2)$ doublet of gravitinos $(\ps_\mu^A,\ps_{\mu A})$
of opposite chirality, where $A=1,2$ is an $SU(2)$ index. The vector multiplet consists of a complex
scalar $z^i$, a vector $A_\mu^i$, where $i=1,2\ldots n_V$ and an $SU(2)$ doublet of gauginos
$(\la^{i A}, \la_A^{\bi})$ with opposite chirality. The hyper multiplets contain scalars $q^X$,
where $X=1,\ldots 4 n_H$ and two hyperinos $(\ze_\al,\ze^\al)$ , $(\al=1\ldots 2 n_H)$ of opposite
chirality. The moduli space of the theory factorizes into a product of a special K\"{a}hler manifold
and a quaternionic K\"{a}hler manifold \cite{Andrianopoli:1996cm}
\beq
\mM = \mS\mK(n_V) \times \mQ(n_H)\ .
\eeq
For a K\"{a}hler manifold the metric is derived from a K\"{a}hler potential
\beq
g_{i\bj}=\p_i\p_{\bj}\mK\ .
\eeq
Since the K\"{a}hler manifold is also special K\"{a}hler there exist local holomorphic
sections $(X^\La,F_\La)$ where $\La=0\ldots n_V$ \footnote{For $\La=0$, $A^0_\mu$ refers to
the graviphoton. }, where  $F_\La=\f{dF}{dX^\La}$ ($F$ is the
holomorphic prepotential). The K\"{a}hler
potential can be expressed in terms of the sections as
\beq
\mK=-\ln(i(\bX^\La F_\La-X^\La\bF_\La))\ .
\eeq
The K\"{a}hler manifold is also symplectic and hence one can introduce symplectic sections
$(L^\La(z,\bz),M_\La(z,\bz))$ that satisfy
\beq
i(\bL^\La M_\La-L^\La \bM_\La)=1\ .
\eeq
The symplectic sections are related to the usual sections via the relations
\beq
(X^\La, F_\La)=e^{-\f{\mK}{2}}(L^\La, M_\La)\ .
\eeq
All the matter couplings in the theory are completely determined in terms of the symplectic
sections. Let $K_\La^i(z)$ be Killing vectors that generate isometries on the manifold $\mS\mK$.
Gauging the symmetries of the manifold amount to replacing ordinary derivatives with gauge
covariant derivatives
\beq\label{sccov}
D_\mu z^i=\p_\mu z^i + K^i_\La A^\La_\mu\ .
\eeq
For the rest of the discussion, the gauge group is abelian for simplicity. 

The hyperscalars $q^X$, $X=1,\ldots 4 n_H$ parametrize the quaternionic K\"{a}hler manifold $\mQ$.
The metric on $\mQ$ is defined by
\beq
ds^2 = g_{XY} dq^X \otimes dq^Y
\eeq
in suitable coordinates $q^X$ on $\mQ$. Since $\mQ$ is also K\"{a}hler the metric can be
derived from a suitable K\"{a}hler potential. The isometries on $\mQ$ are generated by Killing
vectors $K^X_\La$. Once again gauging is done by replacing ordinary derivatives with gauge covariant
derivatives
\begin{align}\label{covder22}
D_\mu q^X =&\p_\mu q^X +A_\mu^\La K_\La^X(q)\ .
\end{align}
In the above we have set the gauge coupling constant to identity for simplicity. Notice that gauging
(\eqref{sccov}, \eqref{covder22}) introduces additional terms in the theory. Supersymmetry
invariance of the resultant action requires the addition of a potential
\beq
\mV(z,\bz,q)= \left((g_{i\bj}K_\La^i K_\Si^{\bj}+4 g_{XY}K_\La^X K_\Si^Y)\bL^\La L^\Si
+(g^{i\bj}f_i^\La f_{\bj}^\Si-3\bL^\La L^\Si) P_\La^x P_\Si^x\right)
\eeq
where $f_i^\La=(\p_i+\f{1}{2}\p_i\mK)L^\La$. The triplet $P_\La^x , x=1,2,3$ are real Killing
prepotentials on the quaternionic K\"{a}hler manifold. The bosonic part of the Lagrangian of the
$\mN=2$ theory takes the form
\begin{align}\label{Lag4}
 \mL=& -\f{1}{2}R + g_{i\bj}D^\mu z^i D_\mu \bz^{\bj}+g_{XY}D_\mu q^X D^\mu q^Y+ 
i(\bN_{\La\Si}\mF^{-\La}_{\mu\nu}\mF^{-\Si\mu\nu}-N_{\La\Si}\mF^{+\La}_{\mu\nu}\mF^{+\Si\mu\nu}
)\nn\\ &+\mV(z,\bz,q)
\end{align}
where $N_{\La\Si}$ are the period matrices.\footnote{These are functions of $z^i$ and can be
expressed in terms of the sections $M_{\La}=N_{\La\Si}L^\Si$.} The self/anti-self dual field
strengths are defined as
\beq
\mF_{\mu\nu}^{\pm \La}=\f{1}{2}\left(F_{\mu\nu}^\La\pm\f{i}{2} \ep_{\mu\nu\rh\si}F^{\La
\rh\si}\right)
\eeq
where the usual field strength is defined as $F^\La_{\mu\nu}=\f{1}{2}(\p_\mu A^\La_\nu-\p_\nu
A^\La_\mu)$. The supersymmetry transformations of the fermionic fields are given by
\begin{align}\label{susy4d}
 \de\ps_{\mu A}=& D_\mu\ep_A + i S_{AB}\ga_\mu \ep^B +2i (\text{Im}N)_{\La\Si} L^\Si
\mF_{\mu\nu}^{-\La}\ga^\nu \ep_{AB}\ep^B\nn\\
\de\la^{iA}=& i D_\mu z^i \ga^\mu \ep^A-g^{i\bj}\bar{f}_{\bj}^\Si
(\text{Im}N)_{\La\Si}\mF_{\mu\nu}^{-\La}\ga^{\mu\nu}\ep^{AB}\ep_B+ W^{iAB}\ep_B\nn\\
\de\ze_\al=&i\mU_X^{B\be}D_\mu q^X \ga^\mu\ep^A \ep_{AB}\ep_{\al\be}+N_\al^A\ep_A
\end{align}
where 
\begin{align}
 S_{AB}=& \f{i}{2} (\si^r)_A^{\ C}\ep_{BC}P_\La^r L^\La\nn\\
 W^{iAB}=& \ep^{AB}k_\La^i \bL^\La +i (\si_r)_C^{\ B}\ep^{CA}P_\La^r g^{i\bj}f_{\bj}^\La\nn\\
 N_\al^A= &2 \mU_{\al X}^A K_\La^X \bL^\La\ .
\end{align}
In the above $\mU_{\al X}^A$ are vielbeins on the quaternionic manifold. The covariant
derivative on the spinor $\ep_A$ is defined as
\beq
D_\mu \ep_A =\na_\mu \ep_A+\f{i}{2}(\si^r)_A^B A_\mu^\La P_\La^r \ep_B
\eeq
where $\na_\mu$ is the covariant derivative defined with respect to the usual spin connection. For
the rest of the discussion we assume a generic gauging of the symmetries of hypermultiplet manifold.

At the attractor point the scalars are independent of spacetime coordinates,
\beq
z^i=\text{const} \ , q^X=\text{const}\ .
\eeq
 The supersymmetry variations \eqref{susy4d} at the attractor point then reduce to
\begin{align}\label{attsusy4d}
 \de\ps_{\mu A}=& D_\mu\ep_A + i S_{AB}\ga_\mu \ep^B +2i (\text{Im}N)_{\La\Si} L^\Si
\mF_{\mu\nu}^{-\La}\ga^\nu \ep_{AB}\ep^B\nn\\
\de\la^{iA}=& -g^{i\bj}\bar{f}_{\bj}^\Si
(\text{Im}N)_{\La\Si}\mF_{\mu\nu}^{-\La}\ga^{\mu\nu}\ep^{AB}\ep_B+ W^{iAB}\ep_B\nn\\
\de\ze_\al=&i\mU_X^{B\be}K^X_\La A_\mu^\La \ga^\mu\ep^A \ep_{AB}\ep_{\al\be}+N_\al^A\ep_A\ .
\end{align}
Setting the gravitino variations to zero, we get the Killing spinor equation
\begin{align}\label{KS4dg}
\p_\mu\ep_A+\f{1}{4}\om_\mu^{\ ab}\ga_{ab}\ep_A+\f{i}{2}(\si_x)_A^B P_\La^x A_\mu^\La\ep_B+ i
S_{AB}\ga_\mu \ep^B +2i (\text{Im}N)_{\La\Si} L^\Si
\mF_{\mu\nu}^{-\La}\ga^\nu \ep_{AB}\ep^B &=0\ .
\end{align}
In the rest of the section, we evaluate the Killing spinor equation \eqref{KS4dg}, the gaugino and
hyperino equations on the background of Bianchi geometries and derive the conditions for
supersymmetry.

\subsection{Bianchi I}\label{4dB1}
Metrics with Bianchi I symmetry in the spatial directions have been studied in the gauged
supergravity literature, the simplest of them being the supersymmetric $AdS_4$ solution
\cite{Ceresole:2001wi}. A supersymmetric Lifshitz solution with exponent $z=2$ has also been
constructed earlier in gauged supergravity by \cite{Halmagyi:2011xh,Cassani:2011sv}. In this section
following the analysis of \cite{Halmagyi:2011xh} we present the supersymmetry conditions
for a simple Bianchi I type - $AdS_2 \times \mathbb{R}^2$ solution. \footnote{
Magnetic $AdS_2\times\mathbb{R}^2$ solutions and their stability have been well explored in the
literature (see for instance \cite{Donos:2011pn,Almuhairi:2011ws,Almheiri:2011cb}).} 

The $AdS_2\times \mathbb{R}^2$  metric has the form
\beq
ds^2= \f{R_0^2}{\si^2}(dt^2-d\si^2)-R_0^2(dy^2+d\rh^2) \ .
\eeq
The Killing vectors along the spatial directions $X_1=\p_y\ , X_2=\p_\rh$ generate the Bianchi I
algebra
\beq
[X_1,X_2]=0\ .
\eeq
 It is easy to construct this metric as a solution to the
equations of motion that follow from the gauged supergravity action \eqref{Lag4}. It is supported
by an electrically charged gauge field whose ansatz we choose to be
\beq\label{gf}
A^{\La}= \f{E^\La}{\si}dt\ .
\eeq
The scalar, gauge field and Einstein equations are listed in \S\ref{AdS2R2eom}. The
Killing spinor equations \eqref{KS4dg} evaluated in the above background are
\begin{align}
\f{\ga^0\si}{R_0}\p_t \ep_A -\f{\ga^{1}}{2R_0}\ep_A+ \f{i G^B_A\ga^0}{2 R_0}\ep_B+i
S_{AB}\ep^B+\f{i N}{2 R_0^2}\ga^{01} \ep_{AB}\ep^B &= 0\label{KS4d21} \\
 \f{\ga^1\si}{R_0}\p_\si \ep_A +i S_{AB}\ep^B+\f{iN}{2 R_0^2}\ga^{01} \ep_{AB}\ep^B &=
0\label{KS4d22}\\
 \f{\ga^2}{R_0}\p_y \ep_A +
i S_{AB}\ep^B-\f{N}{2R_0^2}\ga^{23}\ep_{AB}\ep^B &= 0\label{KS4d23}\\
 \f{\ga^3}{R_0}\p_\rh \ep_A + i S_{AB}\ep^B-\f{N}{2 R_0^2}\ga^{23} \ep_{AB}\ep^B &= 0\label{KS4d24}
\end{align}
where we have defined
\beq\label{defs}
N= (\text{Im} N_{\La\Si})L^\Si E^\La\ , \ G^B_A=(\si_x)_A^{\ B} P_\La^x E^\La\ 
\eeq
for brevity. We choose the following radial ansatz for the Killing spinor
\beq
\ep_A=f(\si)\ch_A
\eeq
where $\ch_A$ is a constant spinor. The difference of \eqref{KS4d22} and
\eqref{KS4d21}  leads to
\beq
\f{\ga^1\si}{R_0}\p_\si\ep_A +\f{\ga^1}{2 R_0}\ep_A- \f{i G_A^B \ga^0}{2 R_0}\ep_B=0\ .
\eeq
The above equation has a simple solution
\beq\label{ansatzads2R22}
f(\si) = \f{1}{\sqrt{\si}}\
\eeq
provided we impose the condition
\beq\label{consp}
E^\La P_\La^x=0\ .
\eeq
We note that this same condition has enabled a supersymmetric Lifshitz solution in 4d
$\mN=2$ gauged supergravity \cite{Halmagyi:2011xh}. Thus the Killing spinor equations reduce to the
algebraic conditions
\begin{align}
 -\f{\ga^1}{2R_0}\ch_A +i S_{AB}\ch^B+\f{iN}{2 R_0^2}\ga^{01} \ep_{AB}\ch^B &= 0 \\
i S_{AB}\ch^B-\f{iN}{2 R_0^2}\ga^{01} \ep_{AB}\ch^B &= 0
\end{align}
where we have substituted $\ga^{23}=-i \ga^{01}\ga_5$ and used $\ga_5\ep^A=-\ep^A$.  It is
straightforward to recast the above equations into the projection conditions 
\begin{align}
\ch_A &=\f{2i N}{R} \ep_{AB}\ga^0 \ch^B \label{AdS2R2proj1}\\
\ch_A &= -4 i R S_{AB}\ga^1 \ch^B \label{AdS2R2proj2}\ .
\end{align}
These projection conditions are very similar to the conditions obtained for the 4d Lifshitz case by
\cite{Halmagyi:2011xh}  (c.f eq 67-68). Squaring  the first projection condition
\eqref{AdS2R2proj1} we get
\beq\label{Ads2R2cons1}
 |N|=\f{R_0}{2} \ .
\eeq
Mutual consistency of the two projectors leads to the equation
\beq\label{AdS2R2proj2f}
\ch_A = 4  R_0 S_{AB}\ga^{10} \ep^{BC}\ch_C\ ,
\eeq
whose self consistency gives the condition
\beq
\sum_{x=1}^3(P^x_\La L^\La)^2= -\f{1}{4 R_0^2}\ .
\eeq
 Note that the triplet of Killing prepotentials $P^x_\La$ are real functions on the
quaternionic manifold. However, the symplectic sections $L^\La$ are 
complex functions in general. For simplicity we can choose the Killing prepotential to lie along the
$x=3$ direction. \footnote{Note that this sets
\beq
P_\La^3 L^\La=\f{i}{2R_0}\ .
\eeq
It is easy to check that this choice is consistent with the projection condition
\eqref{AdS2R2proj2}. Substituting the above in \eqref{AdS2R2proj2} we get
\beq
\ch_A= i (\si_3)_A^{\ D}\ep_{BD}\ga^1\ch^B 
\eeq
that is self consistent.}
Thus the final projection conditions that follow from the gravitino Killing spinor equations are
\begin{align}\label{AdS2R2final}
 \ch_A &=i \ep_{AB}\ga^0 \ch^B \nn\\
\ch_A &=  (\si_3)_A^{\ C}\ga^{10}\ch_C .
\end{align}
These are mutually self consistent projection conditions and together they preserve $\f{1}{4}$ of
the supersymmetry. We now proceed to analyze the gaugino and hyperino conditions in
\eqref{attsusy4d}.

Setting the Hyperino variation \eqref{attsusy4d} to zero, we get the algebraic condition
\beq
i\mU_X^{A\be}K^X_\La \f{E^\La}{R_0} \ga^0\ep^B \ep_{BA}\ep_{\al\be}+2 \mU_{\al X}^A K_\La^X \bL^\La
\ep_A=0\ .
\eeq
We can use the $\f{1}{4}$ BPS projectors \eqref{AdS2R2final} to simplify the above expression to get
\beq
\mU_{X\al}^A K_\La^X\left(\f{E^\La}{R_0}+ 2\bL^\La\right)\ch_A=0\ .
\eeq
An obvious way to solve the condition is to set $E^\La=-2 \bL^\La R_0$. In fact this leads to the
correct equation of motion (second of \eqref{ads2R2eom}). However this leads to an inconsistency
with the known identity (see eq 4.38 of \cite{Andrianopoli:1996cm})
$\text{Im}N_{\La\Si}L^\La\bL^\Si=-\f{1}{2}$ that is true for any $\mN=2$ supergravity. Note that
this was also observed earlier in \cite{Halmagyi:2011xh} for the 4d Lifshitz solution. However we
can solve the hyperino conditions by choosing the Killing vectors to be degenerate on the
quaternionic manifold. In other words, 
\beq
K_\La^X\left(\f{E^\La}{R_0}+ 2\bL^\La\right)=0\ .
\eeq

The gaugino conditions in \eqref{attsusy4d} upon using the $\f{1}{4}$ BPS projections have
the very simple form 
\beq
g^{\i\bj}\bar{f}_{\bj}^\Si\left(
(-\text{Im}N)_{\La\Si}\f{E^\La}{R_0^2}+i P_\Si^3\right)=0\ .
\eeq
This concludes the set of conditions that follow from supersymmetry requirements.
To summarize, the final set of conditions for a $\f{1}{4}$ BPS $AdS_2\times \mathbb{R}^2$ solution
are \begin{align}
& E^\La P_\La^3 =0\ , \ \text{Im}N_{\La\Si}L^\La E^\Si=\f{R_0}{2}\ , \ P_\La^3 L^\La=\f{i}{2R_0}\ ,
\ \nn\\
& K_\La^X\left(\f{E^\La}{R_0}+ 2\bL^\La\right)=0\ , \
g^{\i\bj}\bar{f}_{\bj}^\Si\left(-\text{Im}N_{\La\Si}\f{E^\La}{R_0^2}+i P_\Si^3\right)=0\ .
\end{align}
In addition one has to impose the gauge field equations of motion \eqref{gfeq}. 

\subsection{Bianchi II}\label{4dB2}
In this section we discuss the supersymmetry conditions for a Bianchi II ($AdS_2\times EAdS_2$)
solution of the form
\beq\label{ads2eads21}
ds^2= \f{R_0^2}{\si^2}(dt^2-d\si^2)-\f{R_0^2}{\rh^2}(dy^2+d\rh^2)\ .
\eeq
As discussed in \S\ref{bianchidescription} the symmetries along the spatial directions correspond
to that of $EAdS_2$. Like the previous solution, the $AdS_2
\times EAdS_2$ solution can also be constructed using a time like gauge field \eqref{gf} as source
since it preserves the Bianchi II symmetry along the $(y,\rh$) directions. \footnote{See
\cite{Chimento:2015rra} for magnetic black hole solutions interpolating between $AdS_2 \times
\mathbb{H}^2$ and hyperscale violating solutions at infinity. }
The equations of motion are presented in appendix \S\ref{AdS2EAdS2eom}. The Killing spinor
equations on this background are
\begin{align}\label{ks4d1}
\f{\ga^0\si}{R_0}\p_t \ep_A -\f{\ga^{1}}{2R_0}\ep_A+ \f{i G^B_A\ga^0}{2 R_0}\ep_B+i
S_{AB}\ep^B+\f{i N}{2 R_0^2}\ga^{01} \ep_{AB}\ep^B &= 0\nn\\
 \f{\ga^1\si}{R_0}\p_\si \ep_A +i S_{AB}\ep^B+\f{iN}{2 R_0^2}\ga^{01} \ep_{AB}\ep^B &= 0\nn\\
 \f{\ga^2\rh}{R_0}\p_y \ep_A -\f{\ga^{3}}{2R_0}\ep_A+
i S_{AB}\ep^B-\f{N}{2R_0^2}\ga^{23}\ep_{AB}\ep^B &= 0\nn\\
 \f{\ga^3\rh}{R_0}\p_\rh \ep_A + i S_{AB}\ep^B-\f{N}{2 R_0^2}\ga^{23} \ep_{AB}\ep^B &= 0
\end{align}
where we have defined the quantities $N$ and $G_A^B$ in \eqref{defs}.

Taking the difference of the first and second equations of \eqref{ks4d1}, and similarly the
difference of the third and fourth equations in \eqref{ks4d1} we get the pair of differential
equations
\begin{align}\label{ks4d2}
\f{\si}{R_0} (\ga^0\p_t-\ga^1\p_\si)\ep_A-\f{\ga^1}{2 R_0}\ep_A+\f{i G^B_A\ga^0}{2 R_0}\ep_B &=
0\nn\\
\f{\rh}{R_0}(\ga^2\p_y-\ga^3 \p_\rh)\ep_A -\f{\ga^3}{2 R_0}\ep_A &= 0\ .
\end{align}
Since the $AdS_2\times EAdS_2$ metric factorizes into a product form, with two radii $\rh$ and $\si$
we choose a Killing spinor ansatz of the form
\beq\label{ansatzads2eads2}
\ep_A = \f{1}{\rh^m \si^n}\ch_A
\eeq
where $\ch_A$ is a constant spinor, while $m,n$ take real values. This form of the ansatz is also
consistent with the Bianchi II symmetry of the metric. Substituting the above in \eqref{ks4d2} we
get the conditions
\begin{align}
(2n-1)\ga^1 \ep_A + i G^B_A \ga^0 \ep_B &= 0\nn\\
(2m-1)\ga^3 \ep_A &=0
\end{align}
that can be solved by
\beq\label{condads2eads2}
m=n=\f{1}{2}\ , E^\La P_\La^x=0\ .
\eeq
With the ansatz \eqref{ansatzads2eads2} and the condition \eqref{condads2eads2} the remaining
Killing spinor equations give the conditions
\begin{align}\label{projads2eads2}
\left(\ga^1+\ga^3\right)\ep_A & = 4 i R_0 S_{AB}\ep^B\nn\\
\left(\ga^1-\ga^3\right)\ep_A & = \f{2 i N \ga^{01}}{R_0}\ep_{AB}\ep^B\ .
\end{align}
Unlike the $AdS_2\times \mathbb{R}^2$ case, these conditions are not as simple to work
with. However we can simplify them by multiplying from the left by $\ga^1$ and writing
in terms of the charge conjugate matrix $C=\ga^1\ga^3$ as 
\begin{align}\label{condads2eads2final}
 \left(-1+ C\right)\ch_A &= 4 i R_0 S_{AB}\ga^{01}C(\ch_B)^*\nn\\
 \left(-C+ 1\right)\ch_A &= \f{2 i N}{R_0}\ep_{AB} (\ch_B)^*\nn\\
\end{align}
where we have used $\ch^B=  -\ga_0 C (\ch_B)^*$. We now show that the above condition breaks all
of supersymmetry.  Since $[\ga_5,C]=0$
(see \S \ref{4dspinors}), it is convenient to use a decomposition of the spinor $\ch_A$ in a basis
of simultaneous eigenstates of $\ga_5$ and $C$  as follows
\beq
\ch_A=\left(\begin{tabular}{ c }
        $0$ \\
        $C_A^+|+\ran$ \\
      \end{tabular}\right)+
\left(\begin{tabular}{ c }
        $0$ \\
        $C_A^-|-\ran$ \\
      \end{tabular}\right)
\eeq
where $C_A^+$ and $C_A^-$ are complex coefficients \footnote{Since $(A=1,2)$ there are 8
independent constants in $\ep_A$ as it should be for a $\mN=2$ spinor in four dimension.} and
$|\pm\ran$ are eigenstates of the Pauli matrices. Substituting in the
second equation in \eqref{condads2eads2final} we obtain
\beq
(1-i) C_A^+ |+\ran+ (1+i) C_A^- |-\ran=\f{2 i N}{R_0}\ep_{AB} ((C_B^+)^*|-\ran+(C_B^-)^*|+\ran )\ .
\eeq
Linear independence of the states $|+\ran$ and $|-\ran$ gives rise to the constraints
\begin{align}
 (1-i) C_A^+ &= \f{2i N}{R_0}\ep_{AB}(C_B^-)^*\nn\\
 (1+i) C_A^- &= \f{2i N}{R_0}\ep_{AB}(C_B^+)^*\ .
\end{align}
It is straightforward to see that both of these constraints cannot be satisfied simultaneously as
their mutual consistency leads to
\beq
C_A^+ (1+ \f{2 i |N|^2}{R_0^2})=0\ .
\eeq
Since $|N|^2$ is real, it follows that the only possible solution is that all the $C_A^\pm$
vanish  and hence the metric \eqref{ads2eads21} breaks all the supersymmetry. We now move on to the
five dimensional case where there is a wider variety of solutions with Bianchi symmetries in the
spatial directions.

\section{Bianchi attractors in $\mN=2, d=5$  gauged supergravity}\label{5dgs}
We begin with a brief introduction to $\mN=2, d=5$ gauged supergravity coupled
to $n_V$ vector multiplets with a generic gauging of the very special manifold $\mS$ and the $U(1)_R$ subgroup of the $SU(2)_R$ symmetry group \cite{Ceresole:2000jd}.\footnote{Please note that in all
of five dimensions we use the mostly plus metric signature. }  The very special manifold is parametrized by $n_V+1$ functions $h^I(\ph)$  subject to the constraint
\beq
N\equiv C_{IJK}h^Ih^Jh^K=1
\eeq
where $C_{IJK}$ are constant symmetric tensors. The $\ph^x$, $x=1,2,\ldots n_V$ are scalars in
the $n_V$ vector multiplets. The $I,J$ indices can be raised or lowered using
the ambient metric defined by
\beq
a_{IJ}=-\f{1}{2}\f{\p}{\p h^I}\f{\p}{\p h^J}\ln N\big|_{N=1}\ . 
\eeq
The metric on the moduli space is obtained by pull back of the ambient metric into the moduli space
\beq
g_{xy}= a_{IJ} h^I_x h^J_y \ , \ h^I_x \equiv \f{\p h^I}{\p\ph^x}\ .
\eeq
The Killing vectors $K^I_x$ generate isometries on the very special manifold. These isometries can
be gauged by replacing the ordinary derivatives on the scalars with gauge covariant derivatives
defined by
\begin{align}\label{covder}
D_\mu \ph^x =&\p_\mu \ph^x +g A_\mu^I K_I^x(\ph)
\end{align}
where $A_\mu^I$ are the vectors in the vector multiplet and $g$ is the gauge coupling constant. For the purposes of this paper, we restrict ourselves to the case where the gauge group is abelian.

In addition to gauging the symmetries of the scalar manifold, one can also gauge
the $U(1)_R$ subgroup of the $SU(2)_R$ symmetry of the $\mN=2$ supergravity.\footnote{One can also gauge the $SU(2)_R$ symmetry by including hypermultiplets.} The R symmetry acts as a rotation on
the fermions of the theory and gauging it replaces the usual covariant derivatives on the fermions by gauge
covariant derivatives as
\beq\label{covder3}
D_\mu \ps_{\nu i} = \na_\mu \ps_{\nu i} + g_R A^I_\mu P_{Ii}^{\ \ j}\ps_{\nu j}
\eeq
where $g_R$ is the $U(1)_R$ gauge coupling constant and
\beq\label{U1Rgauging}
P_{Iij}= - V_I \de_{ij} \ .
\eeq
$V_I$ are the Fayet-Illioupoulos parameters. Note that in the above expression $\de_{ij}$ does not play the role of $\ep_{ij}$ as a raising or
lowering operator.  The covariant derivative $\nabla$ is defined with the usual spin connection. Since the
gauging \eqref{covder},and \eqref{covder3} introduces new
terms in the action, supersymmetric closure requires additional terms. These additional terms in
the bosonic part of the Lagrangian take the form of a potential
\beq\label{pot}
\mV(\ph)=-g_R^2 (2 P_{ij}P^{ij}- P^a_{ij}P^{a ij})
\eeq
with the definitions
\begin{align}\label{defss}
 P_{ij}=h^I P_{Iij}\ , \ P^a_{ij}= h^{aI}P_{Iij}\ , \ h^{aI}= f_x^a h^{xI}\ ,
\end{align}
where $f_x^a$ are vielbeins on the very special manifold $\mS$. Note that the potential is unaffected by the gauging of $\mS$. Addition of hypermultiplets and tensor multiplets will change the shape of the potential, however to get $AdS$ vaccum it is sufficient to gauge the $U(1)_R$ symmetry. 

With the various definitions as stated above the bosonic part of the Lagrangian reads as
\cite{Ceresole:2000jd}
\begin{align}\label{lag}
\he^{-1} \mL & = -\f{1}{2}R -\f{1}{4} a_{IJ} F^I_{\mu\nu}F^{J \mu\nu}-\f{1}{2}g_{xy}(\ph)D_\mu\ph^x D^\mu\ph^y
+\f{\he^{-1}}{6\sqrt{6}}C_{IJK} \ep^{\mu\nu\rh\si\ta}F^I_{\mu\nu}F^J_{\rh\si}A^K_\ta -\mV(\ph)
\end{align}
where $\he =\sqrt{-\text{det}g_{\mu\nu} }$. The bosonic sector of the supersymmetry transformations
are
\begin{align}\label{susytransf}
& \de_\ep\ps_{\mu i}= D_\mu\ep_i +\f{i}{4\sqrt{6}} h_IF^{\nu\rh I} (\ga_{\mu\nu\rh}-4
g_{\mu\nu}\ga_\rh)\ep_i +\f{i}{\sqrt{6}}g_R \ga_\mu \ep^jP_{ij}\nn\\
& \de_\ep \la_i^a = -\f{i}{2} f_x^a D_\mu \ph^x \ga^\mu\ep_i +\f{1}{4} h^a_I
F_{\mu\nu}^I\ga^{\mu\nu}\ep_i+g_R\ep^j P_{ij}^a.
\end{align}
The $\la_i^a$ ($i=1,2$ and $a=1,\ldots n_V$) are gauginos in the vector multiplets and $\ep_i$ is a symplectic majorana spinor. The covariant derivative is defined as
\beq
D_\mu\ep_i\equiv\p_\mu\ep_i +\f{1}{4}\om_\mu^{\ ab}\ga_{ab}\ep_i+ g_R A_\mu^I P_{Iij}\ep^j\ .
\eeq
See Appendix \S\ref{5dspinors} for our notations and conventions of 5d gamma matrices.

We are interested in Bianchi type near horizon solutions to \eqref{susytransf} that satisfy attractor conditions. It is well known that at the attractor point the moduli are constants independent of spacetime coordinates
\beq\label{attractorpt}
\ph^x=\text{const} 
\eeq
The field equations that follow from \eqref{lag} are given in \cite{Inbasekar:2012sh}. The
supersymmetry transformations at the attractor point take the form
\begin{align}\label{attrsusytransf}
& \de_\ep\ps_{\mu i}= D_\mu\ep_i +\f{i}{4\sqrt{6}} h_IF^{\nu\rh I} (\ga_{\mu\nu\rh}-4
g_{\mu\nu}\ga_\rh)\ep_i +\f{i}{\sqrt{6}}g_R \ga_\mu \ep^jP_{ij}\nn\\
& \de_\ep \la_i^a = -\f{i}{2}g f_x^a A_\mu^I K^x_I \ga^\mu\ep_i +\f{1}{4} h^a_I
F_{\mu\nu}^I\ga^{\mu\nu}\ep_i+g_R\ep^j P_{ij}^a
\end{align}
In the following sections we evaluate the spinor conditions on the Bianchi attractor backgrounds.
 As discussed in \S\ref{bianchidescription} the Bianchi type metrics have the generic
form\footnote{In this coordinate system, the boundary of the
Poincar\'{e} AdS metric lies at $r\to\infty$.}
\beq
ds^2= \eta_{ab}e^a e^b=L^2\left(-e^{2\be_t r}dt^2+ \et_{ij}(r) \om^i \otimes \om^j+dr^2\right)
\eeq
where $e^a, a=0,\ldots 4,$ are one forms and $L$ is a positive constant that measures the size of
the spacetime. The $\om^i, i=1,\ldots 3$ are one forms manifestly
invariant under the homogeneous symmetries described by the Bianchi classification. 

\subsection{The gaugino conditions}\label{gaugino5d}
In this section, we solve the gaugino conditions
\begin{align}\label{ghvar}
 & \de_\ep \la_i^a = -\f{i}{2}g f_x^a A_\mu^I K^x_I \ga^\mu\ep_i +\f{1}{4} h^a_I
F_{\mu\nu}^I\ga^{\mu\nu}\ep_i-g_R\ep^j h^a_I V^I \de_{ij}=0
\end{align}
where we have substituted \eqref{U1Rgauging} and \eqref{defss}.
In gauged supergravity literature the terms in the supersymmetry variations that are proportional to the
gauge coupling constants are referred to as fermionic shifts. For maximal supersymmetry all of the
fermionic shifts in the gaugino conditions must vanish \cite{Ceresole:2000jd,Inbasekar:2012sh}. From the integrability conditions eq 31 of \cite{Inbasekar:2012sh} it follows that the only maximally supersymmetric Bianchi type solution is
$AdS_5$. Our first result will be to argue that the above result is also true for solutions with matter, in this case the Bianchi type geometries. Then we will require some of the fermionic shifts to vanish and explore conditions for supersymmetric solutions. 

First we focus on the cases when none of the fermionic shifts vanish. Preserving some
amount of supersymmetry from the gaugino and hyperino conditions require that the algebraic
conditions on the constant part of the spinor $\ze_i$ be not too restrictive. In other words, the
matrices that project out the various components of $\ze_i$ must commute with one another. The
projection conditions that can appear on the spinor in the equations \eqref{ghvar} are entirely
dependent on the gauge field configurations. Typically the Bianchi type solutions are sourced by
either timelike or spacelike massive gauge fields and a cosmological constant
\cite{Iizuka:2012iv,Iizuka:2012pn,Inbasekar:2012sh}. (In particular see appendix B of \cite{Iizuka:2012iv} for various choices of gauge field configurations that solve the equations of motion.)
At the attractor point the scalars are constant
and effective mass terms for the gauge fields
\beq\label{massterm}
g^2  K_{IJ}(\phi) A^{I\mu} A^J_\mu
\eeq
appear due to the presence of the gauge covariant derivatives in the supergravity action
\eqref{lag}. Here $K_{IJ}$ is the Killing norm defined as $g_{xy} K^x_I K^y_J$. The mass terms are proportional to the norm of the Killing vectors and to the square of the gauge coupling constant. We analyze two possible cases separately below.

\subsubsection{Non vanishing fermionic shifts}
To begin with we keep our analysis very generic with respect to the gauging of the scalar manifold
(model independent) but specific only to the field content that generates the solution. By this, we mean that there are no specific conditions that the Killing vectors on $\mS$ are required to satisfy. We first
consider the case where the gauge fields have only the time component turned on.
The Bianchi metrics that have been constructed so far \cite{Iizuka:2012iv,Iizuka:2012pn,Inbasekar:2012sh} are
sourced by time like or spacelike gauge fields. Time like gauge fields are of the form
\begin{align}
A&= A(r)dt \nn\\
dA&= \p_r A(r) dr\wedge dt \ .
\end{align}

In order to solve the gaugino conditions \eqref{ghvar} it is necessary to impose
projection conditions on the constant part of the spinor $\ep_i$. From the time like gauge field
configuration it is clear that the following conditions have to be imposed in \eqref{ghvar}
\footnote{The spacetime coordinates are $x^\mu=(t,x^1,x^2,x^3,r)$, while the corresponding tangent
space indices run over $a=0,\ldots 4$.}
\begin{align}\label{tgfp}
 \ga_0\ep_i&=\pm i \ep_i\nn\\
 \ga_{04}\ep_i&=\pm \ep_i\ .
\end{align}
The first projector appears in the $A^\mu\ga_\mu$ terms, while the second appears in the
$F^{\mu\nu}\ga_{\mu\nu}$ terms. While each of the projectors is well defined, it is clear that the
two conditions are mutually incompatible since the projections \eqref{tgfp} are mutually orthogonal.
Thus when the fermionic shifts do not vanish all solutions sourced by time like gauge fields break supersymmetry. 
Thus with a time like gauge field, under gauging it is not possible to obtain supersymmetry preserving
projection conditions. Note that this is completely independent of the functional dependence of the
Killing spinor. 

Let us now consider the case with gauge fields having spacelike components turned on. (For examples, see \cite{Iizuka:2012iv,Iizuka:2012pn,Inbasekar:2012sh})
\begin{align}
A&=A(x,r)\om^i \nn\\
dA&= \p_r A(x,r) dr\wedge \om^i + \p_{x_j} A (x,r) dx^j\wedge \om^i+
\f{1}{2} A(x,r) C_{jk}^{\ \ i} \om^j\wedge \om^k
\end{align}
where $x=x^i, i=1,2,3$ are the directions that have homogeneous symmetries. In this case, it is
easy to see from \eqref{ghvar} that the projections that can appear are
\begin{align}\label{sgfp}
 \gamma_{i}\ep_i&=\pm \ep_i\nn\\
 \gamma_{i4}\ep_i&=\pm i \ep_i\nn\\
 \gamma_{ij}\ep_i&=\pm i \ep_i\ .
\end{align}
In any given configuration for the space like gauge field, the first projector always appears.
Depending on the precise functional dependence the second/third or both second and third projectors
can appear. In any case, we see that the first projector in \eqref{sgfp} is mutually orthogonal to
both the second and third. Thus even with a space like gauge field, under generic conditions it is
not possible to obtain supersymmetry preserving conditions. Note that this too is completely
independent of the functional dependence of the Killing spinor. Thus when the fermionic shifts do not vanish all massive Bianchi attractors are non-supersymmetric in gauged supergravity with generic gauging of the scalar manifold.

For all the solutions in this class, one can study the Killing spinor equations independently and find that the radial spinor breaks supersymmetry. We have summarized these results in appendix \S\ref{masc}. The solutions constructed in \cite{Inbasekar:2012sh} are all of this type and are all non-supersymmetric.

\subsubsection{Vanishing fermionic shifts}
The other possibilities to solve \eqref{ghvar} are situations where some of the fermionic shifts
vanish in special cases. From the studies of the attractor mechanism for black holes in $d=5$
ungauged supergravity, it is known that attractor solutions solve the gaugino conditions
\cite{Larsen:2006xm,Klemm:2000nj} with the extremization of central charge
\beq\label{ms1}
\p_x(Z)=\p_x(h^I Q_I)=0\ , \ h^I V_I= 1\ .
\eeq
Imposing the attractor conditions on  \eqref{ghvar}\footnote{The FI parameters $V_I$ are arbitrary and can be scaled to satisfy this condition.}, we find that the gaugino conditions reduce to
\begin{align}\label{ghvar2}
 & \de_\ep \la_i^a = -\f{i}{2}g f_x^a A_\mu^I K^x_I \ga^\mu\ep_i =0 \ .
\end{align}
Note that the square of this fermionic shift term is proportional to
\beq
g^2 g_{xy} A_\mu^I A^{\mu J } K_I^x K_J^y,
\eeq
the mass term discussed in the introduction of this section. Thus for preserving supersymmetry we
have to set the effective mass term to zero. This can be achieved in two ways. 
\begin{itemize}
\item The trivial choice is $g=0$ or no gauging of the scalar manifold $\mS$. 
\item The other more non-trivial possibility is to find a Killing direction in $\mS$ that satisfies
$K^I_x Q_I=0$ at the attractor point.
\end{itemize}
Note that for the class of models discussed in \cite{Gunaydin:2000xk,Gunaydin:1999zx} studied earlier explicit solutions were constructed and analysed in \cite{Inbasekar:2012sh,Inbasekar:2013vra} and it can be checked that the condition $K^I_xQ_I=0$ is not satisfied. However note that using this condition would kill the effective mass terms in the field equations of motion (see eq 18, eq 22 of \cite{Inbasekar:2012sh}) which is problematic and would only lead to massless solutions.

We pause here to briefly summarize the conclusions of this section. Analyzing the gaugino conditions we have the results that in $\mN=2$ gauged supergravity with a generic gauging of the symmetries
of scalar manifold and a $U(1)$ gauging of the $SU(2)_R$ symmetry,
\begin{itemize}
\item  There are no massive Bianchi attractor solutions that preserve any amount of supersymmetry for a generic gauging when the fermionic shifts do not vanish.  
\item When the extremization condition is met $\p_x(Z) =0$, supersymmetry allows only massless Bianchi solutions. \footnote{This possibility appears to be relaxed when tensor multiplets are included, we comment on this briefly in \S\ref{gens}}
\item For massless Bianchi solutions, the gaugino conditions are completely solved by 
the attractor conditions \eqref{ms1} and there are no additional projection conditions. The amount of
supersymmetry preserved is completely determined by the Killing spinor equations.
\end{itemize}
Some examples of solutions with massless gauge fields are given in appendix \S\ref{new5d}. These
solutions can be easily constructed in Einstein-Maxwell theory with a cosmological constant, actually
all of them can be also constructed easily, for instance in the $U(1)_R$ gauged supergravity model
studied in \cite{Inbasekar:2014lra}.

The last and final possibility for this section corresponds to vacuum solutions in the absence of matter. In this case the gaugino conditions are trivial. The supersymmetry conditions are completely  determined by the Killing spinor equation that follows from the gravitino variation. The solution space includes the well known $AdS_5$ solution
\cite{Ceresole:2000jd,Gunaydin:2000xk,Ceresole:2001wi}, Bianchi III $AdS_3\times
\mathbb{H}^2$ and Bianchi V $AdS_2\times\mathbb{H}^3$ solutions sourced only by a cosmological constant (see appendix \S\ref{new5d}). The results of this section can get modified by addition of tensor and hyper multiplets. We comment on this briefly in \S\ref{gens}.

\subsection{The gravitino conditions: Killing spinor equation}\label{Ksmless}
We have already shown from the gaugino conditions that there are no possible
supersymmetric Bianchi solutions sourced by massive gauge fields when the fermionic shifts do not
vanish for the theory with a generic gauging of scalar manifold and a $U(1)_R$ gauging. For all these cases, one can show by analyzing the Killing spinor equation
that a radial spinor independently gives rise to inconsistent projection conditions. We have summarized the results in \S\ref{masc}. 

In this section we analyze the gravitino Killing spinor equation for the Bianchi solutions sourced
by massless gauge fields. For the $U(1)_R$ gauged supergravity \eqref{U1Rgauging}, the Killing
spinor equation we need to solve is of the form 
\beq\label{ksf}
D_\mu\ep_i +\f{i}{4\sqrt{6}} h_IF^{\nu\rh I} (\ga_{\mu\nu\rh}-4
g_{\mu\nu}\ga_\rh)\ep_i +\f{i}{\sqrt{6}}g_R \ga_\mu \ep_i^{\ k}\ep_k= 0
\eeq
where
\beq
D_\mu\ep_i\equiv\p_\mu\ep_i +\f{1}{4}\om_\mu^{\ ab}\ga_{ab}\ep_i+ g_R A_\mu^I V_I \ep_{i}^{\
k}\ep_k\ 
\eeq
where we have used the attractor conditions \eqref{ms1}. Note that we have used the notation
$\ep^{kj}\de_{ij}=\ep_i^{\ k}$ where $\ep_i^{\ k}$ is numerically same as $-\ep_{ik}$. It follows
that $\ep_i^{\ k}\ep_k^{\ l}= -\de_i^{\ l}$. We need to remember that $\de_{ij}$ is just one
component of the general triplet in $P_{Iij}$ and hence one cannot use $\de_{ij}$ or $\ep_{i}^{\ k}$
to raise or lower the R symmetry index \cite{Gunaydin:1999zx,Ceresole:2000jd}.\footnote{We thank
Antoine Van Proeyen for useful communication regarding this issue.} In the following, we solve the
Killing spinor equations for various Bianchi type geometries.

\subsubsection{Bianchi I $AdS_5$}
As a warm up let us begin our analysis with the simplest known $AdS_5$ metric written in terms of
the one forms
\begin{align}\label{B1formAds}
& e^0= L e^{ r}dt\ , \  e^1= L e^{r} \om^1\ , \ e^2= L e^{ r}\om^2\ , \ e^3= L
e^{ r}\om^3\ , \ e^4= L dr
\end{align}
where $L$ is the $AdS$ scale. The invariant one forms
\beq
\om^i= dx^i \ , i=1,2,3 
\eeq
all commute with one another and satisfy $d\om^i=0$, characteristic of the Bianchi I algebra. Since we are discussing the $U(1)_R$ case, the gaugino conditions are trivial.\footnote{For more general $AdS$ critical points see \cite{Ceresole:2001wi,Gunaydin:2000xk}.}

 The Killing spinor equation \eqref{ksf} in the background \eqref{B1formAds} reads as,
\begin{align}\label{B1ksAds}
 e^{-r}\ga_0 \p_t \ep_i -\f{1}{2} \ga_4\ep_i-\f{i}{\sqrt{6}} L g_R \ep_i^{\ k}\ep_k &=0\nn\\
 e^{- r}\ga_1 \p_{x^1}\ep_i+\f{1}{2}\ga_4\ep_i+\f{i}{\sqrt{6}} L g_R \ep_i^{\ k}\ep_k &=0\nn\\
 e^{- r}\ga_2 \p_{x^2}\ep_i+\f{1}{2}\ga_4\ep_i+\f{i}{\sqrt{6}}L g_R \ep_i^{\ k}\ep_k &=0\nn\\
 e^{-r}\ga_3 \p_{x^3}\ep_i+\f{1}{2}\ga_4\ep_i+\f{i}{\sqrt{6}} L g_R \ep_i^{\ k}\ep_k &=0\nn\\
\ga_4\p_r\ep_i+\f{i}{\sqrt{6}} L g_R \ep_i^{\ k}\ep_k &=0\ .
\end{align}
The following equations can be obtained after some algebraic manipulations
\begin{align}\label{B1ksads1}
 \ga_0 \p_t\ep_i +\ga_a \p_{x^a}\ep_i &=0\nn\\
\ga_a \p_{x^a}\ep_i -\ga_b \p_{x^b}\ep_i &=0\nn\\
 \ga_4 \p_r\ep_i +e^{-r}\ga_0 \p_t\ep_i-\f{1}{2}\ga_4\ep_i&=0\nn\\
 \ga_4 \p_r\ep_i -e^{-r}\ga_a \p_{x^a}\ep_i-\f{1}{2}\ga_4\ep_i&=0
\end{align}
where $a=1,2,3$. There are two independent solutions to the above equations
\begin{align}
\ep_i &= e^{\f{r}{2}}\ze_i^+ \ , \ \ga_4\ze_i^+=\ze_i^+\label{adssol1}\\
\ep_i &=\left( e^{-\f{r}{2}}+e^{\f{r}{2}}(x^m\ga_m)\right)\ze_i^- \ , \ \ga_4\ze_i^-=-\ze_i^-
\label{adssol2}\ .
\end{align}
Each of the spinors \eqref{adssol1} and \eqref{adssol2} preserves $\f{1}{2}$ the supersymmetry and
the full solution enjoys a $\mN=2$ supersymmetry. Substituting the above in \eqref{B1ksAds} we get
the consistency condition
\begin{align}\label{B1ksAds2}
\ze_i^\pm=\mp \f{2i}{\sqrt{6}}L g_R \ep_i^{\ k}\ze_k^\pm\ .
\end{align}
It follows that (note that $\ep_i^{\ k}\ep_k^{\ l}=-\de_i^{\ l}$)
\beq\label{susyAdScondition}
(1-\f{2}{3}L^2 g_R^2 )\ze_i^\pm=0\ .
\eeq
This of course is the equation of motion for $AdS_5$ metric, thus we see that supersymmetry
conditions automatically guarantee the equation of motion. 

\subsubsection{Bianchi I: Anisotropic $AdS_3\times \mathbb{R}^2$}\label{ads3r2}
The anisotropic $AdS_3\times\mathbb{R}^2$ solution can be easily constructed with magnetic fields
and a cosmological constant (see \S \ref{new5d}).\footnote{See for example the isotropic solution
in the $U(1)^3$ truncation of type II supergravity on $S^5$ by \cite{Almuhairi:2011ws}. For general geometries of the type $AdS_3\times\Si_g$ in STU model of supergravity and their dual field theory interpretation see
\cite{Benini:2012cz,Benini:2013cda,Bobev:2014jva,Benini:2015bwz}.} The
metric has the simple form 
\beq\label{ads3r2}
e^0 = e^r dt \ , \ e^1= e^r \om^1\ , \ e^2 = \f{|B|}{2} \om^2\ ,  \ e^3 = \f{|B|}{2}\om^3\ , \ e^4=
dr\ .
\eeq
The magnetic fluxes in the $x^2, x^3$ directions generate anisotropy but preserve the
rotational symmetries of $\mathbb{R}^2$. The solution \eqref{ads3r2} has been of considerable
interest in computations of shear viscosities in anisotropic phases
\cite{Jain:2015txa,Critelli:2014kra}. 
The invariant one forms
\beq
\om^i= dx^i \ , i=1,2,3 
\eeq
all commute with one another and satisfy $d\om^i=0$ of the Bianchi I algebra. In \eqref{ads3r2} 
$|B|=B^I B_I$ is the strength of the magnetic field. We choose our gauge field ansatz such that
\beq
F^I_{x^2 x^3}= B^I\ .
\eeq
The Killing spinor equations in the background are of the form
\begin{align}
 \ga_0 e^{-r} \p_t \ep_i -\f{1}{2}\ga_4\ep_i -\f{i}{\sqrt{6}} \left(\f{Z}{2}\ga_{23}\ep_i+g_R
\ep_i^{\ k}\ep_k\right) &=0 \label{a3r21}\\
 \ga_1 e^{-r} \p_{x^1} \ep_i +\f{1}{2}\ga_4\ep_i +\f{i}{\sqrt{6}} \left(\f{Z}{2}\ga_{23}\ep_i+g_R
\ep_i^{\ k}\ep_k\right) &=0\label{a3r22}\\
 \ga_2 \p_{x^2} \ep_i +\f{i}{\sqrt{6}}\f{|B|}{2} \left(-Z\ga_{23}\ep_i+g_R
\ep_i^{\ k}\ep_k\right) &=0\label{a3r23}\\
 \ga_3 \p_{x^3} \ep_i +\f{i}{\sqrt{6}}\f{|B|}{2} \left(-Z\ga_{23}\ep_i+g_R
\ep_i^{\ k}\ep_k\right) &=0\label{a3r24}\\
 \ga_4 \p_r \ep_i +\f{i}{\sqrt{6}}\left(\f{Z}{2}\ga_{23}\ep_i+g_R
\ep_i^{\ k}\ep_k\right) &=0\label{a3r25}
\end{align}
where $Z=h_I B^I$ is the central charge. In the above we have chosen the following
condition 
\beq
B^I V_I=0\ .
\eeq
This condition is the five dimensional analogue of \eqref{consp}.

It is easy to obtain the following differential equations
from the above set
\begin{align}
 \ga_0 \p_t\ep_i +\ga_1\p_{x^1}\ep_i&=0\nn\\
 \ga_2 \p_{x^2}\ep_i -\ga_3\p_{x^3}\ep_i&=0\nn\\
\ga_4\p_r\ep_i +\ga_0 e^{-r}\p_t \ep_i -\f{1}{2}\ga_4\ep_i &=0\nn\\
\ga_4\p_r\ep_i -\ga_1 e^{-r}\p_{x^1} \ep_i -\f{1}{2}\ga_4\ep_i &=0\ .
\end{align}
Notice the similarity of the above equations to the ones we have obtained in $AdS_5$ case
\eqref{B1ksads1}, except the second equation that suggests that the $x^2,x^3$ directions can scale
differently as compared to the $x^1$ direction. Once again there are two independent solutions to
the above equations
\begin{align}
\ep_i &= e^{\f{r}{2}}\ze_i^+ \ , \ \ga_4\ze_i^+=\ze_i^+\label{ads3r2sol1}\\
\ep_i &=\left( e^{-\f{r}{2}}+e^{\f{r}{2}}(t\ga_0+x^1 \ga_1+\al(x^2\ga_2+x^3\ga_3))\right)\ze_i^- \ ,
\ \ga_4\ze_i^-=-\ze_i^-
\label{ads3r2sol2}\ 
\end{align}
where $\al$ is a real parameter. The projection due to the radial Dirac matrix has the same effect
as in the $AdS$ case, namely the projector preserves one half of the supersymmetry in each of
$\ze^\pm$. Substituting the solution \eqref{ads3r2sol2} in the $x^2,x^3$ equations
\eqref{a3r23}-\eqref{a3r24} we find that $\al=0$. Thus the Killing spinor \eqref{ads3r2sol2} is
independent of the $\mathbb{R}^2$ directions.   

The remaining equations give rise to the conditions
\begin{align}
 \f{1}{2} \ze_i^\pm +\f{i}{\sqrt{6}}\left(\f{Z}{2}\ga_{23}\ze_i^\pm+g_R\ep_i^{\ k}\ze_k^\pm\right)=
&0 \label{a3r2p1}\\
-Z \ga_{23}\ze_i^\pm+g_R\ep_i^{ \ k}\ze_k^\pm=&0\label{a3r2p2}.
\end{align}
It is easy to see that the the above two equations give rise to the conditions
\begin{align}
\ga_{23}\ze_i^\pm &=\ep_i^{\ k}\ze_k^\pm\nn\\
|Z|= |g_R|&=\f{\sqrt{6}}{3}\label{centads3r2}.
\end{align}
The projection above breaks half of the remaining supersymmetries in each of $\ze_\pm$. As a
result, each of $\ze_\pm$ generate $\f{1}{4}$ of the supersymmetry. Thus the solution \eqref{ads3r2}
is a $\f{1}{2}$ BPS solution. 

\subsubsection{Bianchi III and $AdS_3\times \mathbb{H}^2$ }\label{b3}
In this section we construct a superymmetric Bianchi III type solution sourced by a massless
gauge field
\begin{align}\label{B3}
& e^0= L e^{\be r}dt\ , \  e^1= L \om^1 \ , \ e^2= L e^{\be r}\om^2\ , \ e^3=
L \om^3\ , \ e^4= L dr\ ,
\end{align}
where the invariant one forms are
\beq
\om^1= e^{-x^1}dx^2 \ , \om^2=dx^3\ , \om^3= dx^1\ .
\eeq
The spatial part of the metric has the symmetries of $\mathbb{H}^2 \times \mathbb{R}$. The symmetry
algebra due to these Killing vectors form the Bianchi III algebra in the Bianchi classification in
three dimensions. The sub algebra generated by the Killing vectors of $\mathbb{H}^2$ generate the
Bianchi II algebra in two dimensions.

We choose the gauge field to have components along the $\om^1$ direction
\beq
A^I=B^I e^1 .
\eeq
The Killing spinor equations evaluated in the background are ($Z=h_IB^I$)
\begin{align}\label{KSEB3s}
 e^{-\be r}\ga_0\p_t\ep_i -\f{\be}{2} \ga_4\ep_i -\f{i}{\sqrt{6}}\left(\f{1}{2}Z
\ga_{13}\ep_i+L g_R \ep_i^{\ k}\ep_k\right)&=0\\
e^{x^1}\ga_1 \p_{x^2}\ep_i -\f{\ga_3}{2}\ep_i+L g_R B^I V_I \ga_1\ep_i^{\
k}\ep_k+\f{i}{\sqrt{6}}\left(-Z \ga_{13}\ep_i+L g_R \ep_i^{\ k}\ep_k\right)&=0\\
e^{-\be r}\ga_2 \p_{x^3}\ep_i +\f{\be}{2}\ga_4\ep_i+\f{i}{\sqrt{6}}\left(\f{1}{2}Z
\ga_{13}\ep_i+L g_R \ep_i^{\ k}\ep_k\right)&=0\\
\ga_3\p_{x^1}\ep_i +\f{i}{\sqrt{6}}\left(-Z \ga_{13}\ep_i + L g_R \ep_i^{\ k}\ep_k\right)&=0\\
\ga_4\p_r\ep_i +\f{i}{\sqrt{6}}\left(\f{1}{2}Z \ga_{13}\ep_i+L g_R \ep_i^{\ k}\ep_k\right)&=0.
\end{align}
As before we can obtain the following equations from above
\begin{align}\label{B3x}
 \ga_0\p_t\ep_i+\ga_2\p_{x^3}\ep_i&=0\nn\\
e^{-\be r}\ga_0\p_t\ep_i -\f{\be\ga_4}{2}\ep_i+\ga_4\p_r\ep_i &=0\nn\\
e^{-\be r}\ga_2\p_{x^3}\ep_i +\f{\be\ga_4}{2}\ep_i-\ga_4\p_r\ep_i &=0\nn\\
e^{x^1}\ga_{13}\p_{x^2}\ep_i+\p_{x^1}\ep_i+\f{\ep_i}{2}+L g_R B^I V_I \ga_{13}\ep_i^{\
k}\ep_k&=0 .
\end{align}
 The $AdS_3$ part of the Killing spinor will preserve some supersymmetry provided we assume that the
Killing spinor does not depend on the $\mathbb{H}^2$ part. We get the following conditions from the
above set of equations
\begin{align}
\ep_i &= e^{\f{\be r}{2}}\ze_i^+ \ , \ \ga_4\ze_i^+=\ze_i^+\label{ads3h2sol1}\\
\ep_i &=\left( e^{-\f{\be r}{2}}+e^{\f{\be r}{2}}(t\ga_0+x^3 \ga_2)\right)\ze_i^- \ ,
\ \ga_4\ze_i^-=-\ze_i^- \label{ads3h2sol2}\\
\ga_{13}\ze_i^\pm&= \ep_i^{\ k}\ze_k^\pm \ , \ 4L^2 g_R^2 (B_IV^I)^2=1\label{ads3h2sol3} .
\end{align}
As discussed in the previous sections, the two projectors above combine to break half of the total
supersymmetries of the solution. Substituting the above relations in the Killing spinor equation we
find
\begin{align}
 \f{\be}{2} \ze_i^\pm +\f{i}{\sqrt{6}}\left((\f{Z}{2}+L g_R)\ep_i^{\
k}\ze_k^\pm\right)=
&0 \label{a3h2p1}\\
(-Z +L g_R)\ep_i^{ \ k}\ze_k^\pm=&0\label{a3h2p2}\ .
\end{align}
Consistency of the above equations yield the conditions
\beq\label{parads3}
L g_R = Z\ , \ \be =\sqrt{\f{3}{2}} Z\ , \ 4L^2 g_R^2 (B_IV^I)^2=1\ .
\eeq
Thus we have a one parameter family of $\f{1}{2}$ BPS Bianchi III solutions labeled by the central
charge $Z$. 

When the central charge takes the value \eqref{centads3r2} (the one corresponding to the
$AdS_3\times\mathbb{R}^2$ solution), it follows from \eqref{parads3}
that
\beq
L=1\ , \be=1\ .
\eeq
This is the $AdS_3\times\mathbb{H}^2$ solution constructed in \cite{Klemm:2000nj}. 

It is also possible to construct the vacuum $AdS_3\times\mathbb{H}^2$ solution (see
\eqref{metx}). In this case the simplified equations \eqref{B3x} are
\begin{align}\label{B3xvac}
 \ga_0\p_t\ep_i+\ga_2\p_{x^3}\ep_i&=0\nn\\
e^{-r}\ga_0\p_t\ep_i -\f{\ga_4}{2}\ep_i+\ga_4\p_r\ep_i &=0\nn\\
e^{-r}\ga_2\p_{x^3}\ep_i +\f{\ga_4}{2}\ep_i-\ga_4\p_r\ep_i &=0\nn\\
e^{x^1}\ga_{13}\p_{x^2}\ep_i+\f{\ep_i}{2}+\p_{x^1}\ep_i&=0.
\end{align}
Any solution necessarily depends on the $\mathbb{H}^2$  coordinates and breaks
supersymmetry. The $AdS_3$ part of the equations (first three of \eqref{B3xvac}) are solved by the
usual
\begin{align}\label{so1}
 \ep_i =e^{\f{r}{2}}\ze_i^+ \ , \ \ga_4\ze_i^+ = \ze_i^+\nn\\
 \ep_i =(e^{\f{r}{2}}(\ga_0 t+\ga_2 x^3)+e^{-\f{r}{2}})\ze_i^- \ , \ \ga_4\ze_i^- = -\ze_i^-
\end{align}
whereas the $\mathbb{H}^2$ part of the equations (the last equation in \eqref{B3xvac}) are solved by
\begin{align}\label{so2}
 \ep_i &=e^{-\f{x^1}{2}}\ze_i^- \ , \ \ga_3\ze_i^- = -\ze_i^-\nn\\
 \ep_i &=(e^{-\f{x^1}{2}}\ga_1 x^2+e^{\f{x^1}{2} })\ze_i^+ \ , \ \ga_3\ze_i^+ = \ze_i^+ .
\end{align}
We see that $\ze^\pm$ are required to be simultaneous eigenspinors of both $\ga_3$ and $\ga_4$ in
order to solve the full set of equations \eqref{B3xvac}. \footnote{The general solution is a
combination of \eqref{so1} and \eqref{so2}.}
However, that is impossible since the matrices anti commute. Thus the product space in the vacuum
case breaks all supersymmetry. In the charged case, we are able to avoid the spinor being an
eigenspinor of $\ga_3$ due to the condition \eqref{ads3h2sol3}. This is consistent with the
conclusion from the integrability condition eq. 31 of \cite{Inbasekar:2012sh} that $AdS_5$ is the
unique maximally supersymmetric vacuum solution in the theory.

\subsubsection{Bianchi V}
The Bianchi V solution constructed in \cite{Iizuka:2012iv} is of the form
\begin{align}\label{B5form}
& e^0= L e^{\be_t r}dt\ , \  e^1= L \om^1 \ , \ e^2= L \om^2\ , \ e^3= L \om^3\ , \ e^4= L dr
\end{align}
where the invariant one forms are given by
\beq
\om^1= e^{-x^1}dx^2 \ , \om^2=e^{-x^1}dx^3\ , \om^3= dx^1\ .
\eeq
The Bianchi V geometry in this case has the form of $AdS_2\times\mathbb{H} ^3$. The metric is
sourced by a massless time like gauge field
\beq
A^I= E^I e^0\ .
\eeq
The Killing spinor equations in this background take the form ($Z=E^I h_I$)
\begin{align}\label{KSB5}
e^{-\be_t r}\ga_0 \p_t\ep_i- \f{\be_t}{2}\ga_4\ep_i+ g_R L E^I V_I
\ga_0\ep_i^{\ k}\ep_k+\f{i}{\sqrt{6}}\left(\be_t Z \ga_{04}\ep_i
-L g_R \ep_i^{\ k}\ep_k\right) &=0 \nn \\
e^{x^1}\ga_1
\p_{x^2}\ep_i-\f{1}{2}\ga_3\ep_i+\f{i}{\sqrt{6}}\left(\f{\be_t}{2} Z
\ga_{04}\ep_i+Lg_R \ep_i^{\ k}\ep_k\right) &=0\nn\\
e^{x^1}\ga_2\p_{x^3}\ep_i-\f{1}{2}\ga_3\ep_i+\f{i}{\sqrt{6}}\left(\f{\be_t}{2
}
Z \ga_{04}\ep_i+Lg_R \ep_i^{\ k}\ep_k\right) &=0\nn \\
\ga_3\p_{x^1}\ep_i+\f{i}{\sqrt{6}}\left(\f{\be_t}{2} Z \ga_{04}\ep_i+L g_R \ep_i^{\ k}\ep_k
\right)&=0\nn\\
\ga_4 \p_r\ep_i-\f{i}{\sqrt{6}}\left(\be_t Z \ga_{04}\ep_i-L g_R \ep_i^{\ k}\ep_k\right) &=0.
\end{align}
We can write down the following differential equations after some algebraic manipulations
\begin{align}\label{B5x}
e^{-\be_t r} \ga_0\p_t\ep_i+\ga_4\p_r\ep_i -\f{\be_t}{2}\ga_4\ep_i+g_R L E^I V_I
\ga_0\ep_i^{\ k}\ep_k&=0\nn\\
\ga_{1}\p_{x^2}\ep_i-\ga_{2}\p_{x^3}\ep_i&=0\nn\\
e^{x^1}\ga_{13}\p_{x^2}\ep_i+\p_{x^1}\ep_i+\f{\ep_i}{2} &=0\nn\\
e^{x^1}\ga_{23}\p_{x^3}\ep_i+\p_{x^1}\ep_i+\f{\ep_i}{2} &=0.
\end{align}
Following the arguments given in the previous section, we can solve the $AdS_2$ part of the
equations (first in \eqref{B5x}) by
\begin{align}\label{sol5}
 \ep_i &=e^{\f{\be_t r}{2}}\ze_i^+ \ , \ \ga_4\ze_i^+ = \ze_i^+\nn\\
 \ep_i &=(e^{\f{\be_t r}{2}}\ga_0 t+e^{-\f{\be_t r}{2}})\ze_i^- \ , \ \ga_4\ze_i^- = -\ze_i^-
\end{align}
provided we set $E^I V_I = 0$. If $E^IV_I\neq0$ in this case, even the radial spinor breaks all
supersymmetry.\footnote{See \S \ref{B356} for some related details.} Similarly the $\mathbb{H}^3$
part of the equations (last three of \eqref{B5x}) can be solved by
\begin{align}\label{sol6}
 \ep_i &=e^{-\f{x^1}{2}}\ze_i^- \ , \ \ga_3\ze_i^- = -\ze_i^-\nn\\
 \ep_i &=(e^{-\f{x^1}{2}}(\ga_1 x^2+\ga_2 x^3)+e^{\f{x^1}{2} })\ze_i^+ \ , \ \ga_3\ze_i^+ = \ze_i^+.
\end{align}
Once again, we see that the $\ze_\pm$ are required to be simultaneous eigenspinors of $\ga_3$ and
$\ga_4$, that is impossible since the matrices do not commute.\footnote{In this case too, the
general solution of \eqref{B5x} is a combination of \eqref{sol5} and \eqref{sol6}.} Thus the
solution breaks all supersymmetry. The same arguments apply for the vacuum Bianchi V $AdS_2 \times
\mathbb{H}^3$ solution \eqref{metx}.

\subsubsection{Bianchi VII}
The Bianchi VII metric is expressed in terms of the following one forms
\begin{align}\label{B7form}
& e^0= L e^{\be_t r}dt\ , \  e^1= L dx^1\ , \ e^2= L e^{\be r} (\cos(x^1)dx^2 + \sin(x^1) dx^3)\ ,
\ \nn\\
& e^3= L \la e^{\be r} (-\sin(x^1)dx^2 + \cos(x^1) dx^3)\ , \  e^4 = L dr
\end{align}
where $\la$ is a squashing parameter. The gauge field ansatz is of the form
\beq
A^I = B^I e^2
\eeq
where $B^I$ are constants. The Killing spinor equations in the above background take the form
($Z=h_IB^I$)
\begin{align}\label{KS7}
 e^{-\be_t r}\ga_0
\p_t\ep_i-\f{\be_t}{2}\ga_4\ep_i+\f{i}{\sqrt{6}}\left(\f{Z}{2}\left(\be\ga_{24}-\f{\ga_{13}}{\la}
\right)\ep_i- L g_R \ep_i^{\ k}\ep_k\right)&=0\nn\\
\ga_1\p_{x^1}\ep_i-\f{(1+\la^2)}{4\la}\ga_{123}\ep_i-\f{i}{\sqrt{6}}\left(\f{Z}{2}\left(\be\ga_{24}
+\f{2\ga_{13}}{\la}\right)\ep_i- L g_R\ep_i^{\ k}\ep_k\right)&=0\nn\\
\begin{split}
e^{-\be r}\ga_2 (\cos x^1\p_{x^2}+\sin x^1\p_{x^3})\ep_i
+\f{(1-\la^2)}{4\la}\ga_{123}\ep_i+\f{\be}{2}\ga_4\ep_i+L g_R B^IV_I\ga_2\ep_i^{\ k}\ep_k\\
+\f{i}{\sqrt{6}}\left(\f{Z}{2}\left(\f{\ga_{13}}{\la}+2\be\ga_{24}\right)\ep_i+L
g_R \ep_i^{\ k}\ep_k\right)&=0
\end{split}\nn\\
\begin{split}
e^{-\be r}\ga_3 (-\sin x^1\p_{x^2}+\cos x^1\p_{x^3})\ep_i
-\f{(1-\la^2)}{4\la}\ga_{123}\ep_i+\f{\be}{2}\ga_4\ep_i\\
+\f{i}{\sqrt{6}}\left(-\f{Z}{2}\left(\be\ga_{24}+\f{2\ga_{13}}{\la}\right)\ep_i+L
g_R \ep_i^{\ k}\ep_k\right)&=0
\end{split}\nn\\
\ga_4 \p_r \ep_i +\f{i}{\sqrt{6}}\left(\f{Z}{2}\left(\f{\ga_{13}}{\la}+2 \be
\ga_{24}\right)\ep_i+ L g_R \ep_i^{\ k}\ep_k\right) &=0.
\end{align}
After a few algebraic steps we get the following differential equations
\begin{align}\label{B7K2}
 e^{-\be r}\ga_3 (-\sin x^1\p_{x^2}+\cos x^1\p_{x^3})\ep_i
-\ga_1\p_{x^1}\ep_i+\f{\la}{2}\ga_{123}\ep_i+\f{\be}{2}\ga_4\ep_i=&0\nn\\
e^{-\be r}\ga_2 (\cos x^1\p_{x^2}+\sin x^1\p_{x^3})\ep_i+\f{(1-\la^2)}{4\la}\ga_{123}\ep_i-\ga_4
\p_r \ep_i +\f{\be}{2}\ga_4\ep_i+
Lg_R B^IV_I\ga_2\ep_i^{\ k}\ep_k =& 0
\end{align}
that are solved by the radial spinor
\begin{align}\label{B7cond}
&B^I V_I=0\ , \ \ga_{1234}\ep_i=\ep_i \ , \ \be=\la\nn \\
& \ep_i=e^{\f{\left(3\be^2-1\right)}{4\be}r}\ze_i\ .
\end{align}
Substituting \eqref{B7cond} back into \eqref{KS7} we obtain
\begin{align}\label{B7KSf}
 -\f{\be_t}{2}\ga_4\ep_i+\f{i}{\sqrt{6}}\left(\f{Z}{2}\f{(\be^2-1)}{\be}\ga_{24}\ep_i-
L g_R \ep_i^{\ k}\ep_k\right) &= 0 \nn\\
 -\left(\f{1+\be^2}{4\be}\right)\ga_4\ep_i+\f{i}{\sqrt{6}}\left(\f{Z}{2}\f{(\be^2+2)}{\be}\ga_{24}
\ep_i- L g_R \ep_i^{\ k}\ep_k\right) &= 0 \nn\\
 \f{3\be^2-1}{4\be}\ga_4\ep_i+\f{i}{\sqrt{6}}\left(\f{Z}{2}\f{(2\be^2+1)}{\be}\ga_{24}\ep_i+
L g_R \ep_i^{\ k}\ep_k\right) &= 0 .
\end{align}
As a simple check of the equations we see that $\be=1,\be_t=1, Z=0$ correspond to the Bianchi I
($AdS$) solution \eqref{B1ksAds2}. The equations \eqref{B7KSf} lead to the projections
\begin{align}
&\ga_4\ze_i=- i \ep_i^{ \ k}\ze_k \ , \ (\be_t+2\be)^2=6 L^2 g_R^2\nn\\
&\ga_2\ze_i=- i \ze_i \ , \ \left(\f{\be^2-1}{\be^2+1}\right)^2= \f{3 Z^2}{2}\ , \be_t=\f{\be^4+4
\be^2-1}{2\be(1+\be^2)}\ .
\end{align}
It is clear that the additional projection condition due to $\ga_2$ breaks all of the
supersymmetry. Thus the Bianchi VII solution \eqref{B7form} is non supersymmetric. However the
Bianchi VII algebra is a sub algebra of the Poincar\'{e} algebra (see \S\ref{algebrasx}) and hence
also part of the super Poincar\'{e} algebra. It is possible that there are more general solutions in this class that may be supersymmetric.  

\subsection{Including hyper and tensor multiplets}\label{gens}
 In this section, we briefly comment about the possibilities of new supersymmetric solutions due to addition of tensor or hypermultiplets. We will provide formal arguments as explicit solutions such as the ones constructed in \cite{Inbasekar:2012sh} have not been explored yet in specific models with tensor/hypermultiplets. The addition of tensor/hyper multiplets modifies the supersymmetry transformations \eqref{susytransf}. Let us first consider the gravitino equation \cite{Ceresole:2000jd}
\begin{align}\label{attrsusytransf3}
& \de_\ep\ps_{\mu i}= D_\mu\ep_i +\f{i}{4\sqrt{6}} h_{\tilde{M}}\mH^{\tilde{M}\nu\rh} (\ga_{\mu\nu\rh}-4
g_{\mu\nu}\ga_\rh)\ep_i +\f{i}{\sqrt{6}}g_R \ga_\mu \ep^jP_{ij}
\end{align}
where $H^{\tilde{M}}_{\mu\nu}= \{F^I_{\mu\nu}, B^J_{\mu\nu}\}$, $I=0\ldots n_V$ and $J=1,\ldots, n_T$, $B^J_{\mu\nu}$ is an antisymmetric tensor that belongs to the tensor multiplet. The scalars $h_{\tilde{M}}=\{h^I,h^J\}$ are similarly functions of scalars from the vector and tensor multiplets respectively. The addition of hypermultiplets allows more general R symmetry gauging of the full $SU(2)_R$ symmetry group,
\beq
P_{ij}(q) = h^I P_{I ij}(q) = h^I P_I^r(q) (\si_r)_{ij} 
\eeq
where the potentials are now $SU(2)$ valued functions of the hyperscalars in the hypermultiplet. 

First let us consider the case of hypermultiplets turned on, but no tensor multiplets. In this case, the only difference is that the quaternionic prepotential is a $SU(2)$ triplet function of the hyperscalars instead of a singlet for the $U(1)_R$ case. Hence for $\mN=2$ gauged supergravity with $SU(2)_R$ gauging, including vector and hypermultiplets, and a generic gauging of the symmetries of the very special manifold and the quaternionic K\"{a}hler manifold the Killing spinor results that pertain to non-supersymmetric solutions in \S\ref{Ksmless} and \S\ref{masc} continue to hold. \footnote{For the supersymmetric solutions \S\ref{ads3r2} and \S\ref{b3} addition of tensor and hypermultiplets imposes additional new relations from the hyperscalar equations and the tensor field equations of motion. Moreover the parameter space is also enhanced, so one can possibly find new such solutions.  It will be interesting to see if the solutions \S\ref{ads3r2} and \S\ref{b3} continue to remain supersymmetric in suitable models.} Of course, this does not affect the gaugino conditions, but in addition there are new conditions from hyperino equations. We will discuss them shortly. 

With tensor multiplets turned on in addition there are more possibilities. If the tensor fields are oriented carefully there are possibilities of subtle cancellations that can potentially lead to interesting new solutions with supersymmetry preserving projection conditions. However, in the models that have been studied before in \cite{Inbasekar:2012sh} we have not found any such possibility. Nevertheless this requires an independent analysis and it is helpful to obtain some conditions from gaugino and hyperino conditions first to aid in this direction. 

The addition of tensor multiplets also changes the analysis of the gaugino conditions in an interesting way. The gaugino equations acquire an additional term due to tensor multiplets \cite{Ceresole:2000jd}
\begin{align}\label{ghvarx2}
 & \de_\ep \la_i^{\tilde{a}} = -\f{i}{2}g A_\mu^I f_{\tilde{x}}^{\tilde{a}} K^{\tilde{x}}_I \ga^\mu\ep_i +\f{1}{4} h^{\tilde{a}}_{\tilde{M}}
\mH_{\mu\nu}^{\tilde{M}}\ga^{\mu\nu}\ep_i-g_R\ep^j P^{\tilde{a}}_{ij}+g \f{\sqrt{6}}{4} h^I K_I^{\tilde{x}} f_{\tilde{x}}^{\tilde{a}}\ep_i =0
\end{align}
where $\tilde{x}=0, n_V+n_T$ labels the moduli $\phi^{\tilde{x}}$ in the vector and tensor multiplets. The vielbeins $f_{\tilde{x}}^{\tilde{a}}$ live on the tangent space corresponding to the very special manifold $\mS$. If we continue to impose a straightforward generalization of the attractor conditions \eqref{ms1} \footnote{Here $\phi^*$ and $q^*$ are constant attractor values of the moduli and $B^I$ are the tensor charges.}
\beq\label{attth}
\p_{\tilde{x}}((Q_I h^I(\phi^*)+ B_J h^J(\phi^*))=0\ , \ h^I(\phi^*) P_I^{\tilde{x}}(q^*)=1
\eeq
the gaugino equations reduce to
\begin{align}\label{ghvarx3}
 & \de_\ep \la_i^{\tilde{a}} = gf_{\tilde{x}}^{\tilde{a}} K^{\tilde{x}}_I\left(-i  A_\mu^I  \ga^\mu + \f{\sqrt{6}}{2}h^I \right)\ep_i =0
\end{align}
that can be solved for an electric solution by imposing the conditions
\beq\label{cond1g}
\ga^0\ep_i=\pm i \ep_i\ , \ gf_{\tilde{x}}^{\tilde{a}}(\phi^*) K^{\tilde{x}}_I(\phi^*)\left(\pm Q^I + 2\sqrt{6}h^I(\phi^*) \right)=0
\eeq
or by imposing the conditions
\beq\label{dd1}
Q^I K_I^{\tilde{x}}(\phi^*)=0\ , \ h^I(\phi^*) K_I^{\tilde{x}}(\phi^*)=0\ .
\eeq
for either of electric or magnetic solutions. In addition one also has the hyperino conditions \cite{Ceresole:2000jd} at the attractor point
\begin{align}\label{hvar}
 & \de\ze^A = gf_{Xi}^{A} K^{X}_I\left(-i A_\mu^I  \ga^\mu + \f{\sqrt{6}}{2} h^I \right)\ep_i =0
\end{align}
In the above, $K_I^X$ are similarly Killing vectors on the Quaternioni manifold $\mQ$, $f_{XI}^A$ are vielbeins on $\mQ$, g is the gauge coupling constant for the gauging of the symmetries on $\mQ$.
Note that \eqref{hvar} is structurally similar to the gaugino condition \eqref{ghvarx3} after imposing attractor like conditions \eqref{attth}. Thus for electric solutions we can impose
\beq\label{cond1h}
\ga^0\ep_i=\pm i \ep_i\ , \ gf_{\tilde{Xi}}^{A}(q^*) K^{X}_I(q^*)\left(\pm Q^I + 2\sqrt{6}h^I(\phi^*) \right)=0
\eeq
or by imposing the conditions
\beq\label{dd2}
Q^I K_I^{X}(q^*)=0\ , \ h^I(\phi^*) K_I^{X}(q^*)=0\ .
\eeq
for either of electric or magnetic solutions. It is interesting to note that the conditions in \eqref{dd1} \eqref{dd2}namely
\beq\label{flow}
h^I(\phi^*) K_I^{\tilde{x}}(\phi^*)=0\ , \ h^I(\phi^*) K_I^{X}(q^*)=0
\eeq
appear in flow equations that preserve supersymmetry in AdS (see eq 2.60 of \cite{Ceresole:2001wi}). So it seems reasonable to impose the above conditions to find Bianchi attractor solutions that potentially flow to an asymptotic AdS geometry. However the conditions 
\beq
Q^I K_I^{X}(q^*)=0 \ , \ Q^I K_I^{\tilde{x}}(\phi^*)=0
\eeq
are problematic as they kill the effective mass terms in the field equations \cite{Inbasekar:2012sh} and would still lead to massless solutions. Thus one possibility to find more interesting massive Bianchi solutions in the $\mN=2$ theory with vector, tensor and hypermultiplets with generic gauging is to consider solutions sourced by time like gauge fields. Then the gaugino and hyperino equations are satisfied by the attractor condition \eqref{attth}, the projections \eqref{cond1g} and \eqref{cond1h}. However solving the Killing spinor equation would require great care in choosing the tensor field configuration, as we would require a projection condition on the spinor that would commute with that of \eqref{cond1g} and \eqref{cond1h}. We have not found any such solution in the models considered earlier in \cite{Inbasekar:2012sh,Inbasekar:2013vra}. Perhaps instead of trying to find explicit solutions and then verifying supersymmetry it may be useful to analyse the Killing spinor integrability conditions carefully together with the flow conditions \eqref{flow} to determine the possible supersymmetric Bianchi attractor solutions in this theory. We hope to report this in a future work.

\section{Summary}\label{sumry}
In this paper we analyzed the supersymmetry of Bianchi attractors in $\mN=2$ $d=4,5$ 
gauged supergravity. In $d=4$,  we studied the supersymmetry of Bianchi I and II attractors sourced
by electric fields.  In the Bianchi I case, we studied an $AdS_2\times\mathbb{R}^2$ metric sourced
by a time like gauge field. We analyze the gaugino and Killing spinor equations and find that
the radial spinor and its projection condition preserve $1/4$ of the supersymmetry. In the Bianchi
II case, we construct an electric $AdS_2 \times \mathbb{H}^2$ solution and find that the radial
spinor breaks all supersymmetry.\footnote{The magnetic $AdS_2\times\mathbb{H}^2$ is
known to be $\f{1}{8}$ BPS \cite{Chimento:2015rra,Meessen:2012sr}.} The main lesson we learnt from this exercise is that the radial spinor plays an important role in preserving supersymmetry. These results are special cases of the more general analysis of \cite{Halmagyi:2013qoa,Erbin:2014hsa}.

In $d=5$ $\mN=2$ gauged supergravity, we consider the theory with a generic gauging of symmetries of
the scalar manifold and a $U(1)_R$ gauging of the R symmetry. The
Bianchi attractor geometries that can be constructed are sourced by massive or
massless gauge fields. For a generic gauging of the scalar manifold and R symmetry, when the
fermionic shifts in the gaugino and hyperino conditions do not vanish, the projection conditions
that need to be imposed on the Killing spinor depend entirely on the gauge field/field strength
configuration. We show that for the known field configurations that source the Bianchi type
geometries, there are no supersymmetric projections possible. Independently we show that the radial
spinor breaks supersymmetry for all metrics of this class. Thus for a generic gauging of the scalar
manifold and when the fermionic shifts do not vanish there are no supersymmetric Bianchi attractors.  This result for Bianchi type geometries is similar to the result for maximally supersymmetric solutions \cite{Ceresole:2000jd,Ceresole:2001wi}. 

When the central charge of the theory satisfies an extremization condition at the
attractor point \footnote{In the study of the attractor mechanism in $d=5$ ungauged supergravity, It
is well known that central charge satisfies an extremization condition at the attractor points
\cite{Larsen:2006xm}.}
\beq
\p_i Z=0
\eeq
some of the fermionic shifts vanish. Supersymmetry invariance of the resultant equations allow only massless solutions. This prompts the search for Bianchi type metrics sourced by massless gauge fields and cosmological constant. We construct new Bianchi I, Bianchi III, Bianchi V and Bianchi VII classes of solutions sourced by massless gauge fields and a cosmological constant. Since the gaugino conditions are completely solved in these cases, the supersymmetry preserved by the geometries are determined by the Killing spinor equation. In the Bianchi I class we construct an anisotropic $1/2$ BPS $AdS_3\times\mathbb{R}^2$ solution where the anisotropy is generated by a magnetic field. The supersymmetry is entirely due to the $AdS_3$ part and the Killing spinor does not depend on the $\mathbb{R}^2$ directions. We also construct a one parameter family of $1/2$ BPS Bianchi III geometries, labeled by the central charge. When the central charge of the Bianchi III
geometry takes the same value as that of $AdS_3\times\mathbb{R}^2$, the solution reduces to the
known $1/2$ BPS $AdS_3\times\mathbb{H}^2$ solution \cite{Klemm:2000nj}. For the Bianchi V and
Bianchi VII classes the radial spinor breaks all
supersymmetry and hence these are non-supersymmetric geometries. However, the parameters that characterize
these solutions can be chosen in accordance with the stability criterion discussed in
\cite{Inbasekar:2014lra}.

Finally we also construct vacuum Bianchi III ($AdS_3\times\mathbb{H}^2$) and Bianchi V
($AdS_2\times\mathbb{H}^3$) geometries respectively. The solutions for the Killing spinor
equations in both the cases necessarily require dependence on the $\mathbb{H}^2/\mathbb{H}^3$
coordinates respectively. The radial projection matrices for the $AdS$ and the $EAdS$ geometries do
not commute and hence these geometries break all of supersymmetry. This is consistent with the
results from integrability conditions in $\mN=2$ gauged supergravity \cite{Inbasekar:2012sh}. 

In \S\ref{gens} we explored the possibile conditions to find more interesting Bianchi attractor geometries with massive gauge fields. Solutions with time like gauge fields and suitable tensor field configurations may give rise to supersymmetry preserving projection conditions in the Killing spinor equation. However, this has not worked so far in the models considered in \cite{Inbasekar:2012sh,Inbasekar:2013vra}. We hope to explore this more further in future works.

Having constructed some of the simplest supersymmetric Bianchi attractors it is interesting to find
such solutions in theories with more supersymmetry. It will also be interesting to uplift these
solutions to higher dimensional supergravity. The Killing spinor equations suggest that in most
cases if the geometry has an $AdS_n$ part that factorizes, the corresponding Killing spinor is
sufficient to preserve supersymmetry of the whole solution. Having an $AdS$ part may enable the
construction of more general Bianchi attractor geometries. Finally, it will be most interesting to
construct analytic solutions that interpolate to $AdS$. A related issue is the embeddability of the
Bianchi algebra in the Poincar\'{e} or the conformal algebra. The Bianchi I and Bianchi VII algebras
are sub algebras of the Poincar\'{e} algebra, the other Bianchi algebras have scaling type
generators and may presumably be obtained from a truncation of the conformal algebra.

In this work we studied Bianchi attractors in $d=4,5$. Earlier works have constructed Bianchi attractors as generalized attractors in gauged supergravity \cite{Kachru:2011ps,Inbasekar:2012sh,Inbasekar:2013vra}. In the studies of black holes in ungauged supergravity there have been studies on the 4d/5d correspondence where relation between the potential and critical points in $d=4$ and $d=5$ have been elucidated \cite{Ceresole:2007rq}. Similar studies have been performed for gauged supergravity relating black strings in $d=5$ and $AdS_2\times S^2$ in $d=4$ \cite{Hristov:2014eza}. It would be interesting to explore the relation between generalized attractor potenitals in $d=4$ and $d=5$ and their critical points.

\acknowledgments
We would like to thank Prasanta Tripathy and Sandip Trivedi for several stimulating discussions
throughout the course of this work. We would like to thank Arpan Saha, Prasanta Tripathy and Sandip
Trivedi for collaboration during the initial stages of this work. BC would like to thank Amitabh
Virmani for discussions. K.I. would like to thank Antoine Van Proeyen for helpful correspondence. BC
would like to acknowledge the hospitality of I.I.T. Gandhinagar and I.I.T. Kanpur while this work
was in preparation. K.I. would like to acknowledge the hospitality of I.O.P. Bhubaneswar while this
work was in preparation. The work of BC is partly supported by the D.S.T.-Max Planck Partner Group
``Quantum Black Holes'' between I.O.P. Bhubaneswar and A.E.I. Golm. The work of K.I. and R.S. was
supported in part by Infosys Endowment for the study of the Quantum Structure of Space Time.  We
would all like to acknowledge our debt to the people of India for their generous and steady support
for research in basic sciences.

\appendix
\section{Conventions}\label{conv}
\subsection{Gamma matrices and Spinors in four dimensions}\label{4dspinors}
The Clifford algebra in 4 space-time dimensions is 
\beq
\{\ga_a,\ga_b\}=2\eta_{ab}
\eeq
with the metric convention $\eta_{ab}=\{+,-,-,- \}$. \footnote{We follow the conventions of
\cite{Halmagyi:2011xh} for $\mN=2, d=4$ gauged supergravity.} The Dirac matrices in four dimensions
can be chosen to be
\begin{align}\label{gamma_matrices4d}
 \ga^0 &= I_2 \otimes \si_1 \nn\\
 \ga^1 &= i\si_1 \otimes \si_2 \nn \\
 \ga^2 &= i\si_2 \otimes \si_2 \nn\\
 \ga^3 &= i\si_3 \otimes \si_2 
\end{align}
where $\si_i, i=1,2,3$ are the usual Pauli matrices and $I_2$ is the two dimensional unit matrix.
We define the chirality matrix to be $\ga_5=-i\ga_0\ga_1\ga_2\ga_3$ and the charge conjugation
matrix $C=i\ga^2\ga^0=\ga^1\ga^3$. The charge conjugation matrix $C$ has the property
$C^t=-C=C^{-1}$.

In four dimensions we can impose the weyl condition on a four component spinor such that
\begin{align}
& \ga_5 \ep_A=\ep_A\nn\\
& \ga_5 \ep^A=-\ep^A
\end{align}
where the conjugate spinor is defined as
\beq
\ep^A=(\ep_A)^c = \ga_0 C^{-1} (\ep_A)^*=-\ga_0 C (\ep_A)^*\ .
\eeq

We use the following decomposition of the spinors in some sections. Using the fact that
$[\ga_5,C]=0$, we can decompose the spinor into simultaneous eigenstates of $C$
and $\ga_5$ as follows
\beq
\ep_A=\left(\begin{tabular}{ c }
        $0$ \\
        $C_A^+|+\ran$ \\
      \end{tabular}\right)+
\left(\begin{tabular}{ c }
        $0$ \\
        $C_A^-|-\ran$ \\
      \end{tabular}\right)
\eeq
where $C_A^+$ and $C_A^-$ are complex coefficients. The two component states $|+\ran, |-\ran$
\beq
|+\ran=\f{1}{\sqrt{2}}\left(\begin{tabular}{ c }
        $1$ \\
        $i$ \\
      \end{tabular}\right)\ , \
|-\ran=\f{1}{\sqrt{2}}\left(\begin{tabular}{ c }
        $1$ \\
        $-i$ \\
      \end{tabular}\right)
\eeq
are eigenstates of $\sigma$ matrices
\beq
\si^1|\pm\ran=\pm i |\mp\ran\ , \ \si^2|\pm\ran=\pm |\pm\ran\ , \ \si^3|\pm\ran= |\mp\ran\ .
\eeq

\subsection{Gamma matrices and Spinors in five dimensions}\label{5dspinors}
In this section, we summarize our notations and conventions for spinors in five dimensions. We
mostly follow our conventions of \cite{Ceresole:2000jd}. The Clifford algebra in 5 space-time
dimensions is 
\beq
\{\ga_a,\ga_b\}=2\eta_{ab} \ 
\eeq
where the metric signature that is mostly plus. The Dirac matrices in five dimensions are
\begin{align}\label{gamma_matrices}
 \ga^0 &= -i\si_2 \otimes \si_3 \nn\\
 \ga^1 &= -\si_1 \otimes \si_3 \nn \\
 \ga^2 &= I_2 \otimes \si_1 \nn\\
 \ga^3 &= I_2 \otimes \si_2 \nn \\
 \ga^4 &= -i\ga^0\ga^1\ga^2\ga^3=\si_3 \otimes \si_3 
\end{align}
where $\si_i, i=1,2,3$ are the usual Pauli matrices and $I_2$ is the two dimensional unit matrix.
The charge conjugation matrix $C$ has the property $C^t=-C=C^{-1}$ and,
\beq
C \ga^a C^{-1} = (\ga^a)^t 
\eeq
where $C= B \ga^0$, with $B = \ga^3$ such that $B^* B=-1$. The spinors in the theory
carry an $SU(2)$ index which is raised and lowered using $\ep_{ij}$ 
\beq\label{SU2indices}
X^j = \ep^{ji}X_i, \quad X_j=X^i \ep_{ij}
\eeq
with $\ep_{12}=\ep^{12}=1$. 

Spinors in $d=5$ satisfy a symplectic majorana condition. To apply this
condition one needs $B^*B=-1$, even number of Dirac spinors $\psi_i, i=1,\ldots,2n$ and an
antisymmetric real matrix $\Om_{ij}$ with $\Om^2=-1_{2n}$. The symplectic majorana condition on a
generic spinor reads as
\beq\label{symplectic_majorana}
\psi_i^*=\Om_{ij}B\psi_j
\eeq
or equivalently \cite{Ceresole:2000jd} as
\beq
\bar{\psi}^i\equiv (\psi^*_i)^t \ga^0 = (\psi^i)^t C\ .
\eeq
For $\mN =2$ supersymmetry $i=1,2$, and using $\Om_{ij}=\ep_{ij}$ \eqref{symplectic_majorana} reads
as
\beq
\psi^*_1=\ga^3 \psi_2\ .
\eeq
Note that this condition does not reduce the degrees of freedom as compared to a single
unconstrained Dirac spinor. This is because one needs at least a pair of Dirac spinors to apply the
symplectic majorana condition \eqref{symplectic_majorana}. However, it does make the R-symmetry
manifest.

Antisymmetrization of indices in the Dirac matrices is done with the following convention
\beq
\ga_{a_1 a_2 \ldots a_n}= \ga_{[a_1 a_2 \ldots a_n]}= \frac{1}{n!}\sum_{\si \in P_n} Sign(\si)
\ga_{a_{\si(1)}}\ga_{a_{\si(2)}}\ldots \ga_{a_{\si(n)}}\ .
\eeq
In $d=5$ only $I, \ga_a , \ga_{ab}$ form an independent set, other matrices are related by the
general identity for $d=2k+3$
\beq
\ga^{\mu_1\mu_2\ldots\mu_s}=
\frac{-i^{-k+s(s-1)}}{(d-s)!}\ep^{\mu_1\mu_2\ldots\mu_s}\ga_{\mu_{s+1}\ldots\mu_d}\ .
\eeq
We also list some useful identities involving various Dirac matrices \cite{becker2006string},
\begin{align}\label{matrix_identities}
  [\ga_a,\ga_b] =& 2\ga_{ab} \nn \\
  [\ga_h,\ga_{abc}] =& 2\ga_{habc} \nn \\
  [\ga_{abc},\ga_{egh}]= &\eta_{ef}\eta_{gp}\eta_{hk}( 2 \ga_{abc}^{\ \ \ fpk}-36 \de_{[ab}^{ \ \
  [fp}\ga_{c]}^{ \ k]})\ .
\end{align}

\section{Bianchi solutions in 4d gauged supergravity}\label{4ds}
In this section, we list the field equations of the Bianchi I (AdS$_2\times$ ${\mathbb{R}}_2$) and
Bianchi II (AdS$_2\times$ EAdS$_2$) solutions in $\mN=2, d=4$  gauged supergravity. We are
interested in an attractor type solution where the scalars $(z,q)$ are constants independent
of spacetime coordinates and only the hypermultiplets are charged under abelian gauging.
The field equations can be derived from an effective Lagrangian \cite{Halmagyi:2011xh}
\beq
\mL_{eff}=-\f{1}{2}R +\text{Im}N_{\La\Si}F^\La_{\mu\nu}F^{\Si\mu\nu}-\mV(z,\bz,q)+g_{XY}K_\La^X
K_\Si^Y A_\mu^\La A^{\mu\Si}\ .
\eeq
\subsection{Bianchi I: AdS$_2\times$ ${\mathbb{R}}_2$}\label{AdS2R2eom}
We write the AdS$_2\times$ ${\mathbb{R}}_2$ in a convenient coordinate system as
\beq\label{ads2R2}
ds^2= \f{R_0^2}{\si^2}(dt^2-d\si^2)-R_0^2(dy^2+d\rh^2)\ .
\eeq
This metric can be easily supported by an electric gauge field, we choose our gauge field ansatz to
be
\beq
A^{\La}= \f{E^\La}{\si}dt\ .
\eeq
The gauge field equations are 
\beq\label{gfeq}
g_{XY} K_\La^X K_\Si^Y E^\La =0 \ .
\eeq
There are $n_v+1$ equations for the $n_v+1$ variables $E_\La$.  At the attractor point the scalars
are constants, as a result all the spacetime derivatives drop and the
scalar field equations reduce to the extremization of an effective potential (attractor potential)
\begin{align}\label{scalar4d1}
& \f{\p}{\p q^X}\mV_{eff}=0\ , \ \f{\p}{\p z^i}\mV_{eff}=0 \nn\\
& \mV_{eff}= \mV(z,\bz,q)-g_{XY}K_\La^X K_\Si^Y \f{E^\La E^\Si}{R_0^2}+\text{Im}N_{\La\Si} \f{E^\La
E^\Si}{2R_0^4}\ .
\end{align}
There are $n_V$ scalar equations for $z^i$ and $4 n_H$ hyperscalar equations for $q^X$.
 The Einstein equations are
\begin{align}
 0 = & R_0^2\mV_{eff} +2 g_{XY} K^X_\La K^Y_\Si E^\La E^\Si -\text{Im} N_{\La\Si}\f{E^\La
E^\Si}{R_0^2}\nn\\
0 = & -R_0^2 \mV_{eff} + \text{Im} N_{\La\Si}\f{E^\La E^\Si}{R_0^2}\nn\\
-\f{1}{R_0^2}=&\mV_{eff}
\end{align}
where $\mV_{eff}$ is defined in \eqref{scalar4d1}. The above equations can be recast as the
following conditions
\begin{align}\label{ads2R2eom}
\mV(z,\bz,q)=&-\f{1}{2R_0^2}\nn\\
\f{\text{Im}N_{\La\Si}E^\La E^\Si}{R_0^2} =& -1 \nn\\
g_{XY} K_\La^X K_\Si^Y E^\La E^\Si =&0 
\end{align}
to be satisfied for a given specific model.

\subsection{Bianchi II: AdS$_2\times$ EAdS$_2$ solution}\label{AdS2EAdS2eom}
As we have discussed before the Bianchi II symmetries are the isometries of a hyperbolic space
$\mathbb{H}^2$ that is nothing but Euclidean AdS$_2$. In a suitable coordinate system a Bianchi II
metric
takes the following form
\beq\label{ads2eads2}
ds^2= \f{R_1^2}{\si^2}(dt^2-d\si^2)-\f{R_2^2}{\rh^2}(dy^2+d\rh^2)\ .
\eeq
Similar to the AdS$_2\times$ $\mathbb{R}_2$ solution discussed in the earlier section, this metric
can also be supported by electric gauge fields. The gauge field ansatz is identical to the earlier
case.
The scalar field equations and gauge field equations are same as \eqref{scalar4d1} and \eqref{gfeq}
respectively. The Einstein equations take the form
\begin{align}\label{Einst4d}
 -\f{R_1^2}{R_2^2} =& R_1^2\mV_{eff} +2 g_{XY} K_\La^X K_\Si^Y E^\La E^\Si-\text{Im}N_{\La\Si}
\f{E^\La E^\Si}{R_1^2}\nn\\
\f{R_1^2}{R_2^2}=&-R_1^2 \mV_{eff}+\text{Im}N_{\La\Si} \f{E^\La E^\Si}{R_1^2}\nn\\
\mV_{eff}=&-\f{1}{R_1^2}\ .
\end{align}
The above equations can be recast in the form
\begin{align}
\mV(z,\bz,q)=&-\f{1}{2}\left(\f{1}{R_1^2}+\f{1}{R_2^2}\right)\nn\\
\text{Im}N_{\La\Si}E^\La E^\Si =&\f{R_1^2}{R_2^2}(R_1^2-R_2^2)\nn\\
g_{XY} K_\La^X K_\Si^Y E^\La E^\Si=&0\ .
\end{align}
\section{Killing spinor equation for the massive cases}\label{masc}
In \S\ref{gaugino5d} we demonstrated that the gaugino and hyperino conditions break all
supersymmetry for Bianchi type solutions when $g\neq 0$ cases. In this section, we show that the massive solutions studied earlier do not solve the Killing spinor equation for a radial ansatz. The solutions studied in this
section are sourced by time like gauge fields and a cosmological constant. These have been
constructed earlier in \cite{Iizuka:2012iv,Inbasekar:2012sh}. In this section, we show that
(independent of the conclusions from \S\ref{gaugino5d}) a radial Killing spinor \eqref{spinoransatz}
breaks all supersymmetry conditions.  For this section, we just
assume the radial spinor ansatz
\beq\label{spinoransatz}
\ep_i=f(r)\ze_i
\eeq
where $\ze_i$ are constant symplectic majorana spinors. The Killing spinor equation has the form
\beq
D_\mu\ep_i +\f{i}{4\sqrt{6}} h_IF^{\nu\rh I} (\ga_{\mu\nu\rh}-4
g_{\mu\nu}\ga_\rh)\ep_i +\f{i}{\sqrt{6}}g_R \ga_\mu \ep_i^{\ k}\ep_k= 0
\eeq
where
\beq
D_\mu\ep_i\equiv\p_\mu\ep_i +\f{1}{4}\om_\mu^{\ ab}\ga_{ab}\ep_i+ g_R A_\mu^IV_I \ep_i^{\ k}\ep_k\ .
\eeq

\subsection{Bianchi I}\label{5dB1}
We start with a simple Bianchi I type solution (see Appendix C of \cite{Iizuka:2012iv}) sourced by
a magnetic gauge field. The metric is written in terms of the one forms
\begin{align}\label{B1form}
& e^0= L e^{\be r}dt\ , \  e^1= L e^{\be_1 r} \om^1\ , \ e^2= L e^{\be r}\om^2\ , \ e^3= L
e^{\be r}\om^3\ , \ e^4= L dr
\end{align}
where $\be\geq 0,\be_1\geq 0$ are the Lifshitz exponents. The invariant one forms
\beq
\om^i= dx^i \ , i=1,2,3 
\eeq
all commute with one another and satisfy $d\om^i=0$. We choose the gauge field to lie along the
$x^1$ direction
\beq
A^I=B^I e^1
\eeq
where $B^I$ are constants. The Killing spinor equations in the above background have the form
\begin{align}\label{B1ks0}
 e^{-\be r}\ga_0 \p_t \ep_i -\f{\be}{2} \ga_4\ep_i+\f{i}{\sqrt{6}}\left(\f{\be_1}{2} B^I h_I
\ga_{14}\ep_i- L g_R \ep_i^{\ k}\ep_k\right)&=0\nn\\
 e^{-\be_1 r}\ga_1 \p_{x^1}\ep_i+\f{\be_1}{2}\ga_4\ep_i+g_R L B^I \ga_1 V_I \ep_i^{\ k}\ep_k
+\f{i}{\sqrt{6}}\left(\be_1 B^I h_I \ga_{14}\ep_i +L g_R\ep_i^{\ k}\ep_k\right)&=0\nn\\
 e^{-\be r}\ga_2 \p_{x^2}\ep_i+\f{\be}{2}\ga_4\ep_i+\f{i}{\sqrt{6}}\left(-\f{\be_1}{2} B^I h_I
\ga_{14}\ep_i +L g_R \ep_i^{\ k}\ep_k\right)&=0\nn\\
 e^{-\be r}\ga_3 \p_{x^3}\ep_i+\f{\be}{2}\ga_4\ep_i+\f{i}{\sqrt{6}}\left(-\f{\be_1}{2} B^I h_I
\ga_{14}\ep_i +L g_R \ep_i^{\ k}\ep_k\right)&=0\nn\\
\ga_4\p_r\ep_i+\f{i}{\sqrt{6}}\left(\be_1 B^Jh_J \ga_{14}\ep_i+L g_R
\ep_i^{\ k}\ep_k\right)&=0\ .
\end{align}
Using the Killing spinor ansatz \eqref{spinoransatz} we get the following three independent
equations
\begin{align}\label{B1ks}
 -\f{1}{2}\be\ga_4\ep_i+\f{i}{2\sqrt{6}}\be_1 B^J h_J \ga_{14}\ep_i-\f{i}{\sqrt{6}}L g_R
\ep_i^{\ k}\ep_k &=0\nn\\
\f{1}{2}\be_1\ga_4\ep_i+g_RLB^J\ga_1 V_J \ep_i^{\ k} \ep_k+\f{i}{\sqrt{6}}\be_1B^Jh_J\ga_{14}\ep_i+\f{i}{\sqrt
{6}}L g_R
\ep_i^{\ k}\ep_k&=0 \nn\\
\ga_4\p_r\ep_i+\f{i}{\sqrt{6}}\be_1 B^Jh_J \ga_{14}\ep_i+\f{i}{\sqrt{6}}L g_R
\ep_i^{\ k}\ep_k&=0\ .
\end{align}
Taking the difference of the last two equations from the above set we get
\beq\label{B1sol1}
\p_r\ep_i -\f{1}{2}\be_1\ep_i=0 \ \implies \ \ep_i=e^{\f{\be_1}{2}r}\ze_i
\eeq
where we have imposed the condition 
\beq\label{B1sol2}
B^I V_I=0\ .
\eeq
Recollect that this same condition was used earlier for the 4d supersymmetric Lifshitz solution
\cite{Halmagyi:2011xh} and for the $AdS_2\times \mathbb{R}^2$ solution in \S\ref{4dB2}. Substituting
\eqref{B1sol1} and \eqref{B1sol2} in \eqref{B1ks} we get the projection conditions
\begin{align}
 \ga_{14}\ze_i & = X \ep_i^{\ k} \ze_k \label{B1p1}\\
 \ga_4\ze_i &= i Y \ep_i^{\ k}\ze_k \label{B1p2}
\end{align}
where 
\beq
X=\f{2 L g_R (\be-\be_1)}{(\be_1+2\be)\be_1 (B^Jh_J)} \ , \ Y=\f{\sqrt{6}L g_R}{\be_1+2\be}\ .
\eeq
Consistency of \eqref{B1p1} and \eqref{B1p2} as projectors gives the conditions
\begin{align}
(1 - X^2 )\ze_i &=0\label{B1p11}\\
(1 - Y^2 )\ze_i &=0\label{B1p21}\ 
\end{align}
Note that there is no issue with either of the above projectors by themselves as both the
conditions \eqref{B1p11} and \eqref{B1p21} can be individually met.

However mutual consistency of the projectors \eqref{B1p1} and \eqref{B1p2} together gives
\beq
\ga_1\ze_i= -i X Y \ze_i
\eeq
that breaks supersymmetry. More explicitly squaring the projector we see that
\beq\label{B1p1and2}
(1+X^2 Y^2)\ze_i=0 
\eeq
cannot be met. Hence it follows that the only solution to \eqref{B1p1and2} is that all the $\ze_i$ vanish. This implies that the projectors \eqref{B1p1} and \eqref{B1p2} together break all of the supersymmetry.

\subsection{Bianchi II}
The Bianchi II metrics are constructed out of the one forms \cite{Iizuka:2012iv}
\begin{align}\label{B2form}
& e^0= L e^{\be_t r}dt\ , \  e^1= L e^{(\be_2+\be_3) r} \om^1\ , \ e^2= L e^{\be_2 r}\om^2\
,   \ e^3=  L e^{\be_3 r}\om^3\ , \ e^4= L dr
\end{align}
where the invariant one forms $\om^i$ are given by
\beq
\om^1=dx^2-x^1 dx^3\ , \ \om^2=dx^3 \ , \ \om^3=dx^1\ ,
\eeq
and the exponents $\be_i$ are all positive. We choose the gauge field along the time direction
\beq
A^I= E^I e^0\ .
\eeq
The Killing spinor equations in this background with the spinor ansatz \eqref{spinoransatz} are
given by

\begin{align}\label{KSb2}
 e^{-\be_t r}\ga_0\p_t\ep_i -\f{1}{2}\be_t \ga_{4}\ep_i + L g_R E^I
\ga_0 V_I \ep_i^{\ k}\ep_k+\f{i}{\sqrt{6}}\left(h_IE^I \be_t \ga_{04}\ep_i- L g_R\ep_i^{\ k}\ep_k\right) &=0\nn\\
e^{-(\be_2+\be_3)r}\ga_1\p_{x^2}\ep_i+\f{1}{4}\left(2(\be_2+\be_3)\ga_4-\ga_{123}\right)\ep_i
+\f{i}{\sqrt{6}} \left(\f{1}{2}h_IE^I\be_t\ga_{04}\ep_i+ L g_R\ep_i^{\ k}\ep_k\right)&=0\nn\\
e^{-\be_2 r}\ga_2 (\p_{x^3}+x^1 \p_{x^2})\ep_i+\f{1}{4}(\ga_{123}+2\be_2\ga_4)\ep_i
+\f{i}{\sqrt{6}}\left(\f{1}{2}h_IE^I\be_t\ga_{04}\ep_i+ L g_R\ep_i^{\ k}\ep_k\right) &=0\nn\\
e^{-\be_3 r}\ga_3
\p_{x^1}\ep_i+\f{1}{4}(\ga_{123}+2\be_3\ga_4)\ep_i+\f{i}{\sqrt{6}}\left(\f{1}{2}h_I E^I\be_t
\ga_{04}\ep_i+ L g_R \ep_i^{\ k}\ep_k\right)&=0\nn\\
\ga_4\p_r\ep_i-\f{i}{\sqrt{6}}\left(h_I E^I \be_t\ga_{04}\ep_i- L g_R \ep_i^{\ k}\ep_k\right)&=0\ .
\end{align}

\begin{align}
-\f{1}{2}\be_t \ga_{4}\ep_i + g_R E^I
\ga_0V_I\ep_i^{\ k}\ep_k+\f{i}{\sqrt{6}}\left(h_IE^I \be_t \ga_{04}\ep_i- L g_R\ep_i^{\ k}\ep_k\right)
&=0\label{KSb21}\\
\f{1}{4}\left(2(\be_2+\be_3)\ga_4-\ga_{123}\right)\ep_i
+\f{i}{\sqrt{6}} \left(\f{1}{2}h_IE^I\be_t\ga_{04}\ep_i+ L g_R\ep_i^{\ k}\ep_k\right)&=0\label{KSb22}\\
\f{1}{4}(\ga_{123}+2\be_2\ga_4)\ep_i+\f{i}{\sqrt{6}}\left(\f{1}{2}h_IE^I\be_t\ga_{04}\ep_i+ L
g_R\ep_i^{\ k}\ep_k\right) &=0\label{KSb23}\\
\f{1}{4}(\ga_{123}+2\be_3\ga_4)\ep_i+\f{i}{\sqrt{6}}\left(\f{1}{2}h_I E^I\be_t
\ga_{04}\ep_i+ L g_R \ep_i^{\ k}\ep_k\right)&=0\label{KSb24}\\
\ga_4\p_r\ep_i-\f{i}{\sqrt{6}}\left(h_I E^I \be_t\ga_{04}\ep_i- L g_R
\ep_i^{\ k}\ep_k\right)&=0\label{KSb25}\ .
\end{align}
From \eqref{KSb22}, \eqref{KSb23} and \eqref{KSb24} we immediately see that
\begin{align}\label{B2sol2}
\be_2=\be_3&=\be \  \nn\\
(\ga_{1234}+\be)\ze_i&=0\ .
\end{align}
It follows that $\be=+1$.
The equation satisfied by the radial Killing spinor can be determined from \eqref{KSb25} and
\eqref{KSb21}
\beq\label{B2sol1}
\p_r\ep_i -\f{1}{2}\be_t\ep_i=0 \ \implies \ \ep_i=e^{\f{\be_t}{2}r}\ze_i
\eeq
where we have imposed the condition $ E^I P_I^x=0$ as before. The remaining conditions can be
casted in the form of the projection conditions
\begin{align}
 \ga_4\ze_i&=-i Y \ep_i^{\ k}\ze_k\label{B2proj1}\\
 \ga_0 \ze_i &= -i X \ze_i\label{B2proj2}
\end{align}
where
\beq
Y=\f{\sqrt{6}L g_R}{3+\be_t}\ , \ X = \f{\sqrt{6}\be_t h_I E^I}{(3-2\be_t)}\ .
\eeq
The projectors \eqref{B2proj1} and \eqref{B2proj2} imply the following conditions respectively
\begin{align}
(1-X^2)\ze_i=0 &\implies X=\pm 1\\
(1-Y^2 )\ze_i=0 &\implies Y=\pm 1\ .
\end{align}
However mutual consistency of the two projectors gives
\beq
\ga_{04}\ze_i= -X Y \ep_i^{\ k}\ep_k \implies (1+ X^2 Y^2)\ze_i=0
\eeq
that cannot be satisfied due to all the quantities $X,Y$ being real. Hence the only
possible solution is $\ze_i=0$ that breaks all supersymmetry.
\subsection{Bianchi III, V, VI$_h$ }\label{B356}
The Bianchi III ($h=0$), V ($h=1$) and VI ($h\neq0,1$) metrics are constructed out of the forms
\begin{align}\label{B356form}
& e^0= L e^{\be_t r}dt\ , \  e^1= L e^{\be_1 r}\om^1 \ , \ e^2= L e^{\be_2 r}\om^2\ , \ e^3=
L \om^3\ , \ e^4= L dr
\end{align}
where the invariant one forms are given by
\beq
\om^1= e^{-x^1}dx^2 \ , \om^2=e^{-h x^1}dx^3\ , \om^3= dx^1\ .
\eeq
In general the metrics of the Bianchi III, V, VI types are sourced by a time like gauge field
\cite{Iizuka:2012iv,Inbasekar:2013vra}
\beq
A^I= E^I e^0\ .
\eeq
The few special examples that are highly symmetric
We carry out the Killing spinor analysis for a general $h$ and only set it to the
appropriate values when needed. 

\begin{align}
e^{-\be_t r}\ga_0 \p_t\ep_i- \f{\be_t}{2}\ga_4\ep_i+ g_R L E^I 
\ga_0V_I\ep_i^{\ k}\ep_k+\f{i}{\sqrt{6}}\left(\be_t E^I h_I \ga_{04}\ep_i
-L g_R \ep_i^{\ k}\ep_k\right) &=0 \nn \\
e^{-\be_1 r+x^1}\ga_1
\p_{x^2}\ep_i-\f{1}{2}\ga_3\ep_i+\f{\be_1}{2}\ga_4\ep_i+\f{i}{\sqrt{6}}\left(\f{\be_t}{2} E^I h_I
\ga_{04}\ep_i+Lg_R \ep_i^{\ k}\ep_k\right) &=0\nn\\
e^{h x^1-\be_2
r}\ga_2\p_{x^3}\ep_i-\f{h}{2}\ga_3\ep_i+\f{\be_2}{2}\ga_4\ep_i+\f{i}{\sqrt{6}}\left(\f{\be_t}{2}
E^I h_I \ga_{04}\ep_i+Lg_R \ep_i^{\ k}\ep_k\right) &=0\nn \\
\ga_3\p_{x^1}\ep_i+\f{i}{\sqrt{6}}\left(\f{\be_t}{2} E^I h_I \ga_{04}\ep_i+L g_R \ep_i^{\ k}\ep_k
\right)&=0\nn\\
\ga_4 \p_r\ep_i-\f{i}{\sqrt{6}}\left(\be_t E^I h_I \ga_{04}\ep_i-L g_R \ep_i^{\ k}\ep_k\right) &=0
\end{align}

Using the ansatz \eqref{spinoransatz} the Killing spinor equations take the form
\begin{align}
- \f{\be_t}{2}\ga_4\ep_i+ g_R L E^IV_I  \ga_0\ep_i^{\ k}\ep_k+\f{i}{\sqrt{6}}\left(\be_t E^I h_I
\ga_{04}\ep_i-L g_R \ep_i^{\ k}\ep_k\right) &=0 \label{KSB3561} \\
-\f{1}{2}\ga_3\ep_i+\f{\be_1}{2}\ga_4\ep_i+\f{i}{\sqrt{6}}\left(\f{\be_t}{2} E^I h_I
\ga_{04}\ep_i+Lg_R \ep_i^{\ k}\ep_k\right) &=0\label{KSB3562}\\
-\f{1}{2}h\ga_3\ep_i+\f{\be_2}{2}\ga_4\ep_i+\f{i}{\sqrt{6}}\left(\f{\be_t}{2} E^I h_I
\ga_{04}\ep_i+Lg_R \ep_i^{\ k}\ep_k\right) &=0\label{KSB3563} \\
\f{i}{\sqrt{6}}\left(\f{\be_t}{2} E^I h_I \ga_{04}\ep_i+L g_R \ep_i^{\ k}\ep_k\right)&=0\label{KSB3564}\\
\ga_4 \p_r\ep_i-\f{i}{\sqrt{6}}\left(\be_t E^I h_I \ga_{04}\ep_i-L g_R \ep_i^{\ k}\ep_k\right)
&=0\ . \label{KSB3565}
\end{align}
From \eqref{KSB3565} and \eqref{KSB3561} after imposing the condition $E^I V_I=0$ as before, we
get the radial equation
\beq
\p_r\ep_i-\f{1}{2}\be_t \ep_i=0 \implies \ep_i = e^{\f{\be_t}{2}r}\ze_i\ .
\eeq
The remaining equations can be simplified to the following conditions
\begin{align}
 \ga_3\ze_i &=\be_1\ga_4\ze_i \label{KSB35611}\\
 h\ga_3\ze_i &=\be_2\ga_4\ze_i \label{KSB35622}\\
\be_t E^I h_I \ga_{04}\ze_i &= 2 L g_R \ep_i^{\ k}\ze_k\label{KSB35633}\\
\be_t\ga_4\ze_i & = i \sqrt{6} L g_R \ep_i^{\ k}\ze_k \label{KSB35644}\ .
\end{align}
We already see that the condition \eqref{KSB35611} breaks all of the supersymmetry for 
Bianchi III, V and $VI_h$ cases since
\beq
\ga_{34}\ze_i=-\be_1\ze_i \implies (1+\be_1^2)\ze_i=0
\eeq
cannot be satisfied as $\be_1$ has to be real. For Bianchi V $(h=0)$, it is possible to avoid the
equation \eqref{KSB35622} by choosing $\be_2=0$, however the rest of the conditions obviously break
supersymmetry as can be seen below. The remaining conditions from \eqref{KSB35622}-\eqref{KSB35644}
are
\begin{eqnarray}
\left(1+\f{h^2}{\be_2^2}\right)\ze_i=0\label{f3561}\\
 \left(1+ \left(\f{2 Lg_R}{\be_t E^I h_I}\right)^2  \right)\ze_i=0\label{f3562}\\
\left(1-\f{6 L^2 g_R^2}{\be_t^2}\right)\ze_i=0 \label{f3563}\ .
\end{eqnarray}
The last condition \eqref{f3563} in principle can be satisfied for all cases. However the first two
conditions \eqref{f3561}, \eqref{f3562} lead to the solution $\ze_i=0$ and break supersymmetry
explicitly for all of Bianchi III, V and VI$_h$ cases.
\subsection{Bianchi IX}
Our last example is Bianchi IX, the metric is written in terms of the one forms
\begin{align}\label{B9form}
& e^0= L e^{\be_t r}dt\  , \  e^1= L \om^1\  , \ e^2= L \om^2\ , \  e^3= L \la \om^3 \ , \  e^4 = L
dr
\end{align}
where $\la$ is the squashing parameter as in the Bianchi VII case. The one forms invariant under the
Bianchi IX symmetry are
given by 
\begin{align}
\om^1 &= - \sin x^3 dx^1+\sin x^1 \cos x^3 dx^2\nn\\
\om^2 &=  \cos x^3 dx^1+\sin x^1 \sin x^3 dx^2\nn\\
\om^3 &= \cos x^1 dx^2 +dx^3\ .
\end{align}
Following \cite{Kachru:2013voa} we choose the gauge field ansatz to be
\beq
A_1^I = E^I e^0 \ , \ A_2^J = B^J e^3
\eeq
where $E^I$, $B^J$ are constants and $I+J=n_V+1$.
\begin{align}
e^{-\be_t r}\ga_0 \p_t\ep_i -\f{\be_t}{2}\ga_4\ep_i+ L g_R E^I V_I\ga_0 \ep_i^{\ k}\ep_k
 +\f{i}{\sqrt{6}}\biggl( \left(h_I E^I\be_t\ga_{04}-\f{\la h_J B^J}{2}\ga_{12}\right)\ep_i - L g_R
\ep_i^{\ k}\ep_k\biggr)&=0\nn\\
\begin{split}
\ga_1\left(-\sin x^3 \p_{x^1}+\f{\cos x^3}{\sin x^1}\p_{x^2}-\f{\cos x^3}{\tan
x^1}\p_{x^3}\right)\ep_i+\f{\la}{4}\ga_{123}\ep_i+\f{i}{\sqrt{6}}\biggl( \left(\f{h_I
E^I}{2}\be_t\ga_{04}-\la h_J B^J \ga_{12}\right)\ep_i\\
+ L g_R \ep_i^{\ k}\ep_k\biggr) &=0\end{split}\nn\\
\begin{split}
\ga_2\left(\cos x^3 \p_{x^1}+\f{\sin x^3}{\sin x^1}\p_{x^2}-\f{\sin x^3}{\tan
x^1}\p_{x^3}\right)\ep_i+\f{\la}{4}\ga_{123}\ep_i+\f{i}{\sqrt{6}}\biggl( \left(\f{h_I
E^I}{2}\be_t\ga_{04}-\la h_J B^J \ga_{12}\right)\ep_i\\+L g_R
\ep_i^{\ k}\ep_k\biggr) &=0\end{split}\nn\\
 \f{\ga_3}{\la}\p_{x^3}\ep_i +\f{(2-\la^2)}{4\la}\ga_{123}\ep_i + L g_R B^I V_I\ga_3\ep_i^{\ k}\ep_k
+\f{i}{\sqrt{6}}\biggl(\left(\f{h_IE^I}{2}\be_t\ga_{04}+\f{h_JB^J}{2}\la\ga_{12}
\right)\ep_i + L g_R \ep_i^{\ k}\ep_k \biggr)&=0\nn\\
\ga_4 \p_r\ep_i+\f{i}{\sqrt{6}}\left(\left(- h_I E^I \be_t\ga_{04} +\f{h_J B^J}{2}\la
\ga_{12}\right)\ep_i+ L g_R \ep_i^{\ k}\ep_k\right) &=0
\end{align}
Using the ansatz \eqref{spinoransatz} the Killing spinor equations reduce to
\begin{align}
 -\f{\be_t}{2}\ga_4\ep_i+\f{i}{\sqrt{6}}\biggl( \left(h_I
E^I\be_t\ga_{04}-\f{h_J B^J}{2}\la\ga_{12}\right)\ep_i - L g_R
\ep_i^{\ k}\ep_k\biggr)&=0 \label{B9KS1}\\
\f{\la}{4}\ga_{123}\ep_i+\f{i}{\sqrt{6}}\biggl( \left(\f{h_I E^I}{2}\be_t\ga_{04}- h_J B^J
\la\ga_{12}\right)\ep_i+ L g_R \ep_i^{\ k}\ep_k\biggr) &=0\label{B9KS2}\\
\f{(2-\la^2)}{4\la}\ga_{123}\ep_i +\f{i}{\sqrt{6}}\biggl(\left(\f{h_IE^I}{2}\be_t\ga_{04}+\f{h_JB^J}
{2}\la\ga_{12} \right)\ep_i + L g_R \ep_i^{\ k}\ep_k \biggr)&=0 \label{B9KS3}\\
\ga_4 \p_r\ep_i+\f{i}{\sqrt{6}}\left(\left(- h_I E^I \be_t\ga_{04} +\f{h_J B^J}{2}\la
\ga_{12}\right)\ep_i+ L g_R \ep_i^{\ k}\ep_k\right) &=0\label{B9KS4}
\end{align}
where we have set $E^I V_I=0$ and $B^J V_J=0$. The radial equation can be obtained by
adding \eqref{B9KS4} and \eqref{B9KS1}  
\beq
\p_r\ep_i-\f{\be_t}{2}\ep_i=0 \implies \ep_i = e^{\f{\be_t}{2}r}\ze_i\ .
\eeq
Subtracting \eqref{B9KS3} and \eqref{B9KS2} we get the equation
\beq
\f{1-\la^2}{\la}\ga_3\ze_i +\f{3i}{\sqrt{6}}\la h_JB^J \ze_i=0\ .
\eeq
This gives a projector of the type $\ga_3\ze_i = i X \ze_i$ that breaks all supersymmetry. We can
avoid this condition by choosing $\la=1$ (no squashing) and setting the corresponding source field
$B^J=0$. The remaining independent equations are
\begin{align}
 -\f{\be_t}{2} \ga_4\ze_i +\f{i}{\sqrt{6}}\left(h_I E^I \be_t \ga_{04}\ze_i - L g_R
\ep_i^{\ k}\ze_k \right) &=0\label{B9f1}\\
\f{1}{4}\ga_{123}\ze_i +\f{i}{\sqrt{6}}\left(\f{h_I E^I}{2}\be_t\ga_{04}\ze_i + L g_R
\ep_i^{\ k}\ze_k\right) &=0\ . \label{B9f2}
\end{align}
Adding the above equations and using $\ga_{123}=i \ga_{04}$ we get the projection
\beq\label{B9P1}
\ga_0\ze_i = i X \ze_i \ , \ X= \f{1+\sqrt{6} h_I E^I \be_t}{2\be_t}\ .
\eeq
Squaring the above projector it follows that  $X=\pm 1$. We can solve for $\be_t$ to get 
\beq
\be_t = \pm \f{1}{2\mp\sqrt{6}E^I h_I}\ .
\eeq
Substituting \eqref{B9P1} in \eqref{B9f1} and \eqref{B9f2} we obtain
\beq\label{B9P2}
\ga_4\ze_i=i Y \ep_i^{\ k}\ze_k \  , \ Y= \f{\sqrt{6}Lg_R}{\be_t\pm 1}\ .
\eeq
Squaring the above projector we get the condition
\beq
(1- Y^2)\ze_i =0
\eeq
that constrains the parameters in $Y$. However as we have already seen in the previous cases we find
that the projectors \eqref{B9P1} and \eqref{B9P2} together break all supersymmetry. Acting with
\eqref{B9P1} on \eqref{B9P2} we get
\beq
\ga_{04}\ze_i=- XY \ep_i^{\ k}\ze_k \implies \left(1+ X^2 Y^2\right)\ze_i=0
\eeq
that forces $\ze_i=0$.

\section{Bianchi solutions in 5d Einstein Maxwell theory with a cosmological constant}\label{new5d}
In this section, we list the Bianchi type solutions sourced by a massless gauge field and a
cosmological constant. The action of the system  we consider is  
\begin{equation}\label{action}
 S = \int d^5 x \sqrt{-g} \left\lbrace R + \Lambda - {1 \over 4} F^2   \right\rbrace
\end{equation}
 We note that in our conventions $\Lambda>0$ corresponds to $AdS$ space. The Einstein equations
read as
\beq
\label{eequations}
R^{\mu}_{\nu}-{1\over 2} \delta^\mu_\nu R= T^\mu_\nu.
\eeq
with
\beq
T^\mu_\nu={1\over2}F^{\mu}_{\ \lambda} F_\nu^{\ \lambda} + {1\over2}\delta^\mu_\nu\left(
\Lambda-{1\over4}F^{\rho \sigma} F_{\rho \sigma}\right)
\eeq
We list the details of the various solutions below. All of the below solutions can also be
constructed in the $U(1)_R$ gauged supergravity model considered in \cite{Inbasekar:2014lra}.
\subsection{Bianchi I : $AdS_3\times\mathbb{R}^2$}
The metric is  
\begin{align}
\begin{split} 
ds^2= -  e^{2 r} dt^{2} + dr^{2} + e^{2 r} dx^{2}+ c dy^{2} + c dz^{2}. \\
\end{split}
\end{align}
where c is an undetermined constant. We choose the magnetic field along the $\mathbb{R}^2$ direction
\beq
\label{gf1}
F_{yz}=B,
\eeq
with $B$ being  a constant. It is easy to see that there are two independent equations
\begin{eqnarray}
B^2 - 2 c^2 (-2 +\Lambda) &=& 0  \\
 \frac{B^2}{2 c} - \frac{1}{2}c (-6 +\Lambda) &=& 0
\end{eqnarray}
that are solved by
$$c=\f{|B|}{2}\  , \ \La=4 .$$

\subsection{Bianchi III  $AdS_3\times\mathbb{H}^2$ and Bianchi V $AdS_2\times \mathbb{H}^3$}
The $AdS_3\times\mathbb{H}^2$ solutions have been constructed earlier with massless gauge
fields \cite{Inbasekar:2014lra}. Here we present a vacuum solution as well
\begin{align}
\begin{split} \label{met3}
ds^2= -  e^{2 r} dt^{2} + dr^{2} + e^{2 r} dx^{2}+  dy^{2} + e^{-2 \lambda y} dz^{2}. \\
\end{split}
\end{align}
It is straightforward to  check that the Einstein equations are solved by 
$$\lambda=\sqrt{2}\ ,\  \Lambda=6. $$ Similarly, there is a vacuum $AdS_2\times\mathbb{H}^3$
solution similar to the charged solution constructed in \cite{Iizuka:2012iv}. It is given by 
\begin{align}
\begin{split} \label{metx}
ds^2= -  e^{2 \beta_t r} dt^{2} + dr^{2} +  dx^{2}+  e^{-2 x} dy^{2} + e^{-2 x} dz^{2}. \\
\end{split}
\end{align}
In this case too, the Einstein equations are solved by $\beta_t=\sqrt{2}, \Lambda=6$.

\subsection{Bianchi VII}
The Bianchi VII metric is given by
\beq
\label{metricd}
ds^2=R^2[dr^2 -e^{2\beta_t r} dt^2+ (dx^1)^2 + e^{2\beta r} ((\omega^2)^2 + \lambda^2
(\omega^3)^2)].
\eeq
where the invariant one forms are defined as
\begin{equation}\label{defineomega}
 \omega^1= dx^1 \ ; \ \omega^2=\cos(x^1) dx^2 + \sin(x^1) dx^3 \ ; \ \omega^3=- \sin(x^1)dx^2
+ \cos(x^1)dx^3.
\end{equation}
The gauge field configuration is
\beq
\label{fgp2}
A=e^{\beta r}\left(\sqrt{\tilde{A_2}} \ \omega^2 \right).
\eeq
It follows that the gauge field equations of motion are
\begin{equation}
\label{eomb}
 \lambda^2 (-2 \beta(\beta+\beta_t))+2 = 0.
\end{equation}
Note that the solution we seek has five parameters, $R, \beta_t,\beta, \lambda,$ which enter
in the metric and $\tilde{A}_2$, that determines the gauge field. These are all constants. The
Einstein equations long the $tt$, $rr$, and $x^1x^1$  directions read as
\begin{eqnarray}
\f{2(1+\tilde{ A}_2)}{\lambda^2}+2 \lambda^2+\tilde{A}_2 (2 \beta^2)+24 \beta^2
-4(1+\Lambda) &=& 0\\
 \f{2(1+\tilde{ A}_2)}{\lambda^2}+2 \lambda^2+\tilde{ A}_2 (-2 \beta^2)+8 \beta
(\beta+2\beta_t)-4(1+\Lambda) &=& 0\\
\f{2(1+\tilde{ A}_2)}{\lambda^2}+2 \lambda^2-\tilde{ A}_2 (2 \beta^2)-8(3\beta^2+2
\beta_t \beta+\beta_t^2) -4(1-\Lambda) &=& 0.
\end{eqnarray}
The components  along the $x^2,x^3$ directions lead to 
\begin{eqnarray}
\f{2(3+\tilde{ A}_2)}{\lambda^2}-2 \lambda^2-\tilde{ A}_2 (2
\beta^2)+8(\beta^2+\beta_t \beta+\beta_t^2) -4(1+\Lambda) &=& 0\\
\f{2(1+\tilde{ A}_2)}{\lambda^2}-6\lambda^2-\tilde{ A}_2 (2 \beta^2)-8(\beta^2+\beta_t
\beta+\beta_t^2) +4(1+\Lambda) &=& 0.
\end{eqnarray}
 
Counting eq.(\ref{eomb}) these are $6$ equations in all.  One can check that only $5$ of these are
independent. 
These $5$ equations determine the $5$ parameters $(\Lambda, \beta,\beta_t, \lambda, \tilde{A}_2)$,
which then 
completely determines the solution. 
 Solving for the 5 parameters gives :
\begin{eqnarray}
\label{valpara}
\lambda &=& 1.31964\\
\Lambda &=& 3.28755\\
\beta &=& 0.40893 \\
\beta_t &=& 0.995299
 \label{valparabetat}\\
\tilde{ A}_2&=& 5.73563
\end{eqnarray}

\section{Bianchi algebras from Poincar\'{e} algebra}\label{algebrasx}
In this section, we illustrate that the Bianchi I and VII algebras are easily embedded in the
Poincar\'{e} algebra. We begin with the Conformal algebra in a d dimensional spacetime 
\begin{align}
 [ P_\mu, P_\nu ] &=0 \ \nn\\
[P_\rh, L_{\mu\nu}] &= i \left(\eta_{\mu\rh}P_\nu-\eta_{\nu\rh}P_\mu\right)\nn\\
[L_{\mu\nu}, L_{\rh\si}] &= i
\left(\eta_{\nu\rh}L_{\mu\si}+\eta_{\mu\si}L_{\nu\rh}-\et_{\mu\rh}L_{\nu\si}-\et_{\nu\si}L_{\mu\rh}
\right)\nn\\
[D,P_\mu]& =iP_\mu\nn\\
[D,K_\mu]&=-i K_\mu\nn\\
[K_\mu,P_\nu]&=2 i\left(\eta_{\mu\nu}D- L_{\mu\nu}\right)\nn\\
[K_\rh,L_{\mu\nu}]&=i\left(\eta_{\rh\mu}K_\nu-\eta_{\rh\nu}K_\mu\right)\ .
\end{align}
The first three algebras form the Poincar\'{e} sub algebra of the conformal algebra. 

Consider scaling the
coordinates $$(\la_1x^1, \la_2 x^2, \la_3 x^3) .$$
As a result the generators scale as
\begin{align}
 &P_{i}\to \f{1}{\la_i}  P_{i}\nn\\
& L_{i j}\to \f{1}{\la_i\la_j}L_{ij}\nn\\
& D\to D\nn\\
&\et_{ij}\to \f{1}{\la_i\la_j} \et_{ij}\nn\\
& K_i\to \f{1}{\la_i}K_i
\end{align}

The Bianchi I is generated by the usual translations
\begin{align}
\p_1&= P_1\nn\\
\p_2&= P_2\nn\\
\p_3&= P_3
\end{align}
and the Bianchi $VII$ generators are the combination of translation and rotation generators
\begin{align}
 \p_2 &= P_2\nn\\
\p_3&= P_3\nn\\
\p_1 - x_3\p_2+ x_2 \p_3&=P_1+ L_{23}
\end{align}
Thus the Bianchi I and Bianchi VII algebras form a closed sub algebra of the Poincar\'{e} algebra.
It follows that the algebras are also sub algebras of the super Poincar\'{e} algebra. The other
Bianchi type algebras listed in \cite{ryan1975homogeneous,1975classical,Iizuka:2012iv} are not
embedded in the Poincar\'{e} algebra in any obvious way. 
\bibliographystyle{ieeetr}
\bibliography{sbpap}

\end{document}